\documentclass[12pt,amssymb]{article} 

\usepackage{mathrsfs}
\usepackage{amssymb}
\usepackage[dvips]{graphicx}

\newcommand{\newsection}[1]{
\addtocounter{section}{1} \setcounter{equation}{0}
\setcounter{subsection}{0} \addcontentsline{toc}{section}{\protect
\numberline{\arabic{section}}{{\rm #1}}} \vglue .6cm \pagebreak[3]
\noindent{ \bf  \thesection. #1}\nopagebreak[4]\par\vskip .3cm}
\newcommand{\newsubsection}[1]{
\addtocounter{subsection}{1}\setcounter{subsubsection}{0}
\addcontentsline{toc}{subsection}{\protect
\numberline{\arabic{section}.\arabic{subsection}}{#1}} \vglue .4cm
\pagebreak[3] \noindent{\it \thesubsection.
#1}\nopagebreak[4]\par\vskip .3cm}

\makeatletter
\newcommand{\seclabel}[1]{%
  \@bsphack
  \protected@write\@auxout{}%
     {\string\newlabel{#1}{{\thesection}{\thepage}}}
  \@esphack
  }
\newcommand{\subseclabel}[1]{%
  \@bsphack
  \protected@write\@auxout{}%
     {\string\newlabel{#1}{{\thesubsection}{\thepage}}}
  \@esphack
  }
\newcommand{\tablabel}[1]{%
  \@bsphack
  \protected@write\@auxout{}%
     {\string\newlabel{#1}{{\arabic{tabnum}}{\thepage}}}
  \@esphack
  }
\makeatother

\renewcommand{\theequation}{\thesection.\arabic{equation}}

\newlength{\extraspace}
\newlength{\extraspaces}
\newcounter{dummy}
\newcommand{\bc}{\begin{center}}
\newcommand{\ec}{\end{center}}
\newcommand{\be}{\begin{equation}
\addtolength{\abovedisplayskip}{\extraspaces}
\addtolength{\belowdisplayskip}{\extraspaces}
\addtolength{\abovedisplayshortskip}{\extraspace}
\addtolength{\belowdisplayshortskip}{\extraspace}}
\newcommand{\ee}{\end{equation}}

\newcommand{\ba}{\begin{eqnarray}
\addtolength{\abovedisplayskip}{\extraspaces}
\addtolength{\belowdisplayskip}{\extraspaces}
\addtolength{\abovedisplayshortskip}{\extraspace}
\addtolength{\belowdisplayshortskip}{\extraspace}}
\newcommand{\ea}{\end{eqnarray}}

\newcommand{\ban}{\begin{eqnarray*}
\addtolength{\abovedisplayskip}{\extraspaces}
\addtolength{\belowdisplayskip}{\extraspaces}
\addtolength{\abovedisplayshortskip}{\extraspace}
\addtolength{\belowdisplayshortskip}{\extraspace}}
\newcommand{\ean}{\end{eqnarray*}}
\newcommand{\baa}{
\addtocounter{equation}{1} \setcounter{dummy}{\value{equation}}
\setcounter{equation}{0}
\renewcommand{\theequation}{\thesection.\arabic{dummy}\alph{equation}}
\begin{eqnarray}
\addtolength{\abovedisplayskip}{\extraspaces}
\addtolength{\belowdisplayskip}{\extraspaces}
\addtolength{\abovedisplayshortskip}{\extraspace}
\addtolength{\belowdisplayshortskip}{\extraspace}}
\newcommand{\eaa}{
\end{eqnarray}
\setcounter{equation}{\value{dummy}}
\renewcommand{\theequation}{\thesection.\arabic{equation}}}

\input epsf

\newcounter{fignum}
\newcounter{tabel}

\newcounter{tabnum}
\newcommand{\vev}[1]{\left\langle #1\right\rangle}

\newcommand{\half}{\frac{1}{2}}
\newcommand{\del}{\partial}
\newcommand{\delb}{\bar{\del}}
\newcommand{\eol}{\nonumber \\}

\newcommand{\cO}{{\cal O}}

\newcommand{\Hom}{{\rm Hom}}

\newcommand{\Gr}{{\rm Gr}}
\newcommand{\W}{{\rm W}}

\begin{document}

%
%

\begin{flushright}
December 2012\\
\end{flushright}
\vspace{2cm}

\thispagestyle{empty}

\begin{center}
{\Large\bf The Sen Limit
 \\[13mm] }

{\sc Adrian Clingher}\\[2.5mm]
{\it Department of Mathematics, University of Missouri - St. Louis, USA  \\
\& Institute of Mathematics of the Romanian Academy}\\[5mm]

{\sc Ron Donagi}\\[2.5mm]
{\it Department of Mathematics, University of Pennsylvania \\
Philadelphia, PA 19104-6395, USA}\\[5mm]

{\sc Martijn Wijnholt}\\[2.5mm]
{\it Arnold Sommerfeld Center, Ludwig-Maximilians Universit\"at\\
Theresienstrasse 37 \\
D-80333 M\"unchen, Germany }\\
[15mm]

Abstract:
\end{center}

$F$-theory compactifications on elliptic Calabi-Yau manifolds may
be related to IIb compactifications by taking a certain limit in
complex structure moduli space, introduced by A. Sen. The limit has
been characterized on the basis of $SL(2,{\bf Z})$ monodromies of
the elliptic fibration. Instead, we introduce a stable version of
the Sen limit. In this picture the elliptic Calabi-Yau splits into
two pieces, a ${\bf P}^1$-bundle and a conic bundle, and the
intersection yields the IIb space-time. We get a precise match
between $F$-theory and perturbative type IIb. The correspondence is
holographic, in the sense that physical quantities seemingly spread
in the bulk of the $F$-theory Calabi-Yau may be rewritten as
expressions on the log boundary. Smoothing the $F$-theory
Calabi-Yau corresponds to summing up the $D({-1})$-instanton
corrections to the IIb theory.

\newpage

\renewcommand{\Large}{\normalsize}

\tableofcontents

\newpage

\newsection{Introduction}

$F$-theory was introduced to study vacua with $7$-branes and
varying axio-dilaton \cite{Vafa:1996xn}. Its most common
definition (which however covers only one branch of the moduli
space \cite{Donagi:2011jy}) is obtained by considering a weakly
coupled $M$-theory compactification on an elliptically fibered
Calabi-Yau $\pi:Y \to B$, and taking the limit as the area of the
elliptic fibers shrinks to zero. The axio-dilaton of type IIb
supergravity on $B$ is identified with the modular parameter of
the elliptic fiber. The main advantage of this point of view is
that while the axio-dilaton is a complicated multi-valued function
over $B$, the geometric description $\pi:Y \to B$ allows for a
global description without branch cuts.

Since $F$-theory was originally thought of as a strong coupling
generalization of type IIb, it was natural to look for a precise
limit of the elliptically fibered Calabi-Yau in which one should
recover the perturbative IIb theory. Such a limit was proposed by
A.~Sen, and is now commonly referred to as the Sen limit of an
$F$-theory compactification. In a nutshell, Sen's idea is the
following. Since the elliptic Calabi-Yau has a section, it can be
put in Weierstrass form
\be
y^2 \ = \ x^3 + fx + g
\ee
The modular parameter $\tau$ is identified with the varying
axio-dilaton $i e^{-\phi} + a$ of type IIb. The $j$-function of
the elliptic fiber can now be expressed as
\be j(\tau) \ = \ {4(24f)^3 \over \Delta}, \qquad \Delta = 4 f^3 +
27 g^2 \ee
To recover perturbative IIb, we want $g_s \to 0$, so we want $\tau
\to i \infty$ except possibly at the location of the $7$-branes.
The most generic way to do this is as follows. We express the
Weierstrass coefficients as
\be
f \sim b_2^3 + \cO(\epsilon), \qquad g \sim b_2^2 + \cO(\epsilon)
\ee
By picking suitable coefficients, the leading terms cancel and we
get $\Delta \propto \epsilon$. Then as $\epsilon \to 0$ we have
$\tau \to i \infty$ (and hence $g_s \to 0$) except at $b_2=0$. The
locus $b_2=0$ on $B$ is eventually identified with the orientifold
locus in type IIb. The $D7$-brane locus depends on the
$\cO(\epsilon)$ terms.

Although this perspective has led to interesting applications, it
leaves a number of issues unaddressed. One of the main problems is
that the precise mapping between the $F$-theory data and the IIb
data has never been established. For example, part of the
$F$-theory data is the specification of a configuration for a
three-form field ${\sf C}_3$. Qualitatively, it has been
understood that this should yield the $7$-brane gauge fields and
two-form tensor fields in the IIb limit, but the precise
dictionary was never found. Similarly, there were problems in the
comparison of tadpole constraints, instantons effects and other
things.

We believe that these difficulties indicate that the usual method
of analyzing the $SL(2,{\bf Z})$ monodromy representation really
isn't the right way to think about the Sen limit. We may ask the
question, what are the right tools to address this problem?

Some recent progress was obtained in \cite{Donagi:2010pd}. Instead
of focusing on the monodromy representation, the idea was to look
in more detail at the degenerate $F$-theory Calabi-Yau appearing
in the Sen limit. It was found for example that the IIb Calabi-Yau
$X_{n-1}$, whose appearance looks somewhat mysterious in Sen's
approach, emerged naturally as a certain divisor of singularities
in the limit. It was also found that differential forms with
logarithmic singularities play an important role in the
comparison. However the picture in \cite{Donagi:2010pd} was still
too singular to establish a complete dictionary, particularly for
the relation between the $7$-brane gauge fields in IIb and the
$F$-theory three-form.

In the present paper, we continue this line of thought. As we have
explained in more detail elsewhere \cite{DKW}, we can get a good
dictionary if we construct a stable version of the degeneration.
So in the present paper we will introduce a stable version of the
Sen limit.

Finding a stable version turns out to be remarkably easy, as
generically we only need a single blow-up of Sen's family. One
finds that the elliptic Calabi-Yau splits into two pieces, a ${\bf
P}^1$-bundle and a conic bundle. When applied to elliptic
$K3$-surfaces, this is the $SO(32)$ picture of
\cite{Aspinwall:1997ye,Clingher:2003ui}, as expected from the
$SO(32)$ heterotic/type I/IIb orientifold duality chain in eight
dimensions. For a more general $F$-theory compactification, the Sen
limit is a generalization of the $SO(32)$ limit.

The picture that emerges is that the $D7/O7$ configuration obtained
by Sen is very similar to spectral data of $SO(2n)$ type, and the
dictionary between $F$-theory and the $D7$ data of type IIb takes
the form of a cylinder mapping, even when there is no
$K3$-fibration. This allows us to get a complete map between the
holomorphic data in $F$-theory and type IIb. Furthermore, the
picture we obtain is now completely analogous to the one for the
$E_8\times E_8$ degeneration \cite{Curio:1998bva}, and fits
beautifully with the general picture for weak coupling limits
advocated in \cite{DKW}. As for the $E_8\times E_8$ degeneration or
the general picture in \cite{DKW}, the $SL(2,{\bf Z})$ monodromy
representation plays no role in the Hodge theoretic approach.

Our stable family provides a solid basis for understanding the Sen
limit. In \cite{DKW} we analyzed the limiting mixed Hodge structure
of a certain class of degenerations, generalizing the work of
\cite{Clingher:2003ui}. The Sen limit belongs precisely to the
class of degenerations considered in \cite{DKW}, so in section
\ref{Periods} we specialize the analysis of \cite{DKW} to this
case, and recover the expected form of the IIb action.

The nilpotent orbit theorem further shows that the corrections due
to smoothing the $F$-theory Calabi-Yau have the characteristic
form of $D(-1)$-instanton corrections to perturbative IIb. Initial
evidence for this interpretation of the corrections computed by
$F$-theory was given by Sen and in \cite{Banks:1996nj}, where the
$D(-1)$-instantons were related to the instantons of
Seiberg-Witten theory. There has been recent progress on computing
these corrections directly using localization techniques in the
IIb theory \cite{Fucito:2011kb}.

\newpage

\newsection{The IIb limit as a stable degeneration}

\newsubsection{Sen's description of the limit}

Let us start with some generalities. The data of an $F$-theory
compactification consists of an elliptically fibered Calabi-Yau
manifold  $\pi: Y_n \to B_{n-1}$ with section, and a configuration
for a three-form field ${\sf C}_3$ with flux ${\sf G}_4 = d{\sf
C}_3$. In the present section, the focus will be on aspects of the
Calabi-Yau geometry. The elliptic fibration can be represented in
Weierstrass form:
\be
y^2 \ = \ x^3 + f x + g
\ee
In order to fulfill the Calabi-Yau condition, $f$ and $g$ must be
sections of $K_B^{-4}$ and $K_B^{-6}$ respectively. In addition, we
have to specify a suitable Deligne cohomology class, whose discrete
part yields the ${\sf G}$-flux. This part of the data will be
ignored in the next two subsections.

In the physics literature, the Sen limit is specified as follows
\cite{Sen:1996vd,Sen:1997gv}. We parametrize the Weierstrass
fibration as
\ba\label{fginb} f &=& -{1\over 48}( {\sf b}_2^2 - 24 {\sf b}_4) \eol
g &=& -{1\over 864} (-{\sf b}_2^3 + 36 {\sf b}_2 {\sf b}_4 -216 {\sf b}_6)
\ea
for some choices of the ${\sf b}_i$, which are sections of
$K_B^{-i}$. The coefficients are slightly different from Sen's, and
were chosen so as to emphasize the relation to the ${\sf b}_i$
appearing in Tate's algorithm. Now we introduce a parameter $t$ as
follows:
\ba f &=& -{1\over 48}( {\sf b}_2^2 - 24 t\,{\sf b}_4) \eol g &=&
-{1\over 864} (-{\sf b}_2^3 + 36 t\,{\sf b}_2 {\sf b}_4 -216t^2\,
{\sf b}_6) \ea
The discriminant is given by
\ba \Delta &=& t^2(-{\sf b}_2^2 {\sf b}_8 -8t\,{\sf b}_4^3 -27t^2
{\sf b}_6^2 + 9 t\,{\sf b}_2 {\sf b}_4 {\sf b}_6) \eol [1mm]
 &\sim &   -{1\over 4} t^2 \, {\sf b}_2^2({\sf b}_2 {\sf b}_6 -{\sf b}_4^2) + \cO(t^3) \ea
Using the approximation $j(\tau) \sim \exp(-2\pi i\tau)$ for large
${\rm Im}(\tau)$, we see that
\be \exp(-2\pi i \tau) \ \sim\ {{\sf b}_2^4\over t^2({\sf b}_2 {\sf
b}_6 -{\sf b}_4^2)} , \qquad \tau\ =\ {i\over g_s} + a\ee
in the $t \to 0$ limit. Thus the IIb string coupling goes to zero
almost everywhere, except possibly at ${\sf b}_2 = 0$, and we may
expect a weakly coupled IIb vacuum.

The axion is still multi-valued. In the $t \to 0$ limit all the
roots of the discriminant are located at ${\sf b}_2=0$ and ${\sf
b}_2 {\sf b}_6 -{\sf b}_4^2=0$. In order to relate this to IIb
data, one looks at the $t \to 0$ limit of the $SL(2,{\bf Z})$
monodromy representation
\be \rho: \pi_1(B\backslash \Delta, pt) \ \to \ SL(2,{\bf Z}) \ee
where $pt$ is a base point. The monodromies around these roots were
analyzed in \cite{Sen:1997gv,Sen:1996vd}, with the result that
\be
 {\sf b}_2=0: \left(
                \begin{array}{cc}
                  -1 & 4 \\
                  0 & -1 \\
                \end{array}
              \right)
 , \qquad {\sf b}_2 {\sf b}_6 -{\sf b}_4^2=0:
 \left(
   \begin{array}{cc}
     1 & 1 \\
     0 & 1 \\
   \end{array}
 \right)
  \ee
In the type IIb theory, these $SL(2,{\bf Z})$ monodromies are
generated by $O7$ and $D7$ planes respectively, so this means that
we should interpret the components of the discriminant locus at $t
= 0$ as follows:
\be
 O7:{\sf b}_2=0, \qquad D7: {\sf b}_2 {\sf b}_6 -{\sf b}_4^2=0 \ee
Therefore we get the following picture \cite{Sen:1997gv}: since
${\sf b}_2=0$ is the orientifold locus, the emerging $X_{n-1}$ is
simply the double cover over $B_{n-1}$ with branch locus given by
${\sf b}_2=0$, obtained by undoing the orientifold projection. That
is, in the limit of complex structure moduli space that we discussed
above, the Calabi-Yau manifold $Y_n$ gives rise to a Calabi-Yau
$(n-1)$-fold $X_{n-1}$ given by
\be \xi^2\ =\ {\sf b}_2 \ee
where ${\sf b}_2 \sim K_{B_{n-1}}^{-2}, \xi \sim
K_{B_{n-1}}^{-1}$. The orientifold involution is given by
\be \xi \to -\xi\ee
and the positions of the branes on this $(n-1)$-fold are given as
above. The $D7$ locus on $X_{n-1}$ is simply the pre-image of ${\sf
b}_2 {\sf b}_6 -{\sf b}_4^2=0$ in $B_{n-1}$ under the orientifold
projection.

The Sen limit has received significant attention recently, see for
example \cite{Aluffi:2007sx,Collinucci:2008pf,Collinucci:2008zs,
Donagi:2009ra,Aluffi:2009tm,Esole:2011cn,Esole:2012tf}.

In this approach, the appearance of $X_{n-1}$ looks somewhat
mysterious, and it is not clear how physical quantities in
$F$-theory are related to physical quantities on $X_{n-1}$ with
$7$-branes. For example in compactifications to four dimensions, we
would like to know the relation between the $4d$ superpotentials
computed by $F$-theory and perturbative IIb. It is impossible to
establish such relations with the methods above. We now turn to a
different approach, which will allow us to derive such relations.

\newsubsection{Stable version of the Sen degeneration}

Let us examine the limit of the elliptic Calabi-Yau more closely.
With a little bit of algebra, one finds that we can rewrite Sen's
family of Weierstrass fibrations in the following suggestive form
\be
 y^2\ =\ {1\over 1728}\, [3 {\sf b}_2 - s] s^2 - {{\sf b}_4\over 24}\, t s + {{\sf b}_6\over 4}\, t^2
\ee
Here we defined the new variable
\be
s\ \equiv\ {\sf b}_2 - 12 x
\ee
We consider the family as an $n+1$ fold ${\cal Y}_{n+1}$, together
with a projection $\pi_{\cal Y}:{\cal Y} \to {\sf D}$, where ${\sf
D}$ is the disk parametrized by $t$.

As it stands, the degeneration above is too severe to extract all
the relevant information. For example, we see that if we set $t
=0$, then ${\sf b}_4$ and ${\sf b}_6$ drop out of the equation. As
a result, information about the $D7$-branes appears to be lost.

As we have discussed in detail in \cite{DKW}, we can recover this
information if we instead consider a semi-stable version of the
degeneration. The family above does not provide a semi-stable
degeneration. The variety ${\cal Y}_{n+1}$ is clearly not smooth
as an $(n+1)$-fold and the central fiber $Y_0 = \pi_{\cal
Y}^{-1}(0)$ has singularities worse than normal crossing. We can
fix this by blowing up the family to resolve the singularities.

Our $(n+1)$-fold ${\cal Y}_{n+1}$ has conic singularities along the
sublocus given by $y =s = t=0$, which further degenerate when ${\sf
b}_2 = 0$. In the generic situation, we can desingularize by doing
a single blow-up of ${\cal Y}_{n+1}$. This produces a new family
$\pi_{\widetilde {\cal Y}}:{\widetilde {\cal Y}}_{n+1} \to {\sf
D}$. The effect of the blow-up is to replace central fiber $Y_0$ of
the old family ${\cal Y}$ by its proper transform and the
exceptional divisor of the blow-up. The only remaining
singularities of the central fiber are of normal crossing type,
which is practically as good as a smooth variety. Thus after the
blow-up, we do have a semi-stable degeneration, in fact a stable
one.

Then over $t = 0$, we get a new Calabi-Yau $n$-fold
\be \widetilde Y_0\ =\ W_T \cup_{X_{n-1}} W_E \ee
where $W_T$ is the proper transform of the original fiber at $t
=0$, and $W_E$ is the exceptional divisor created by the blow-up.
In the generic situation, both $W_T$ and $W_E$ are smooth, and no
additional blow-ups are necessary. As we will explain below, the
normal crossing divisor
\be
X_{n-1} \ = \ W_T \cap W_E
\ee
is a double cover of $B_{n-1}$ and should be identified with the
IIb Calabi-Yau. Further, the geometry of lines on $W_E$ encodes
the $D7$-branes.

The original fiber at $t = 0$ is given by
\be y^2\ =\  {1\over 1728}\, [3 {\sf b}_2 - s] s^2 \ =\
{1\over 864}({\sf b}_2 +6x) ({\sf b}_2 - 12 x)^2 \ee
Introducing a new coordinate $\tilde y = y/s$, we
can write this as
\be\label{LineEq} \tilde y^2 \ = \ {1\over 864}({\sf b}_2 +6x) \ee
This is the equation of a rational curve. The map $(x,\tilde y) \to
(x,y)$ identifies the two points
\be (x,\tilde y) \ = \ (-{\sf b}_2/12,\,\pm \sqrt{-{\sf b}_2/576})
\ee
on each fiber. Over ${\sf b}_2 = 0$, the elliptic fiber degenerates
to a cusp. The proper transform $W_T$ replaces the double points by
two distinct points, with monodromy around ${\sf b}_2 = 0$. As
pointed out in \cite{Donagi:2010pd}, these two points fibered over
$B_{n-1}$ give precisely the Calabi-Yau $(n-1)$-fold $X_{n-1}$ which
Sen identified as the IIb space-time, before orientifolding. Indeed,
$X_{n-1}$ is an anti-canonical divisor of $W_T$, so it is
automatically Calabi-Yau, and should be thought of as the `boundary'
of $W_T$. So it is natural to identify this with the IIb space-time.
The ${\bf Z}_2$ involution used for orientifolding exchanges the two
sheets and the $O7$-planes are by definition located at the fixed
points of this involution, which is given by ${\sf b}_2 = 0$.

Now we discuss the geometry of the exceptional divisor $W_E$. It
consists of a fibration of conics over $B_{n-1}$:
\be
\label{ConicEq} y^2 \ = \  {3{\sf b}_2\over 1728} \, u^2 - {{\sf b}_4 \over 24} uv + {{\sf b}_6\over 4} v^2
\ee
We write this as
\be
y^2 = \vec{u}^{\,T}\,
Q \,\vec{u}, \qquad
\vec{u}\ = \
\left(
  \begin{array}{c}
    u \\
    v \\
  \end{array}
\right), \qquad
Q \ = \ {1\over 576}\left(
          \begin{array}{cc}
            {{\sf b}_2} & 12\,{{\sf b}_4 } \\
            12\,{{\sf b}_4} & 144\,{{\sf b}_6} \\
          \end{array}
        \right)
\ee
The discriminant of this conic bundle
is given by
\be
\Delta_{W_E} \ =\ \det(Q) \ = \ {1\over 4}({\sf b}_2 {\sf b}_6 - {\sf b}_4^2) \ = \ 0
\ee
Over the discriminant locus, the quadratic form on the right-hand
side factorizes. Thus the generic fiber of $W_E$ is a ${\bf CP}^1$,
but over $\Delta_{W_E} \ = \ 0$ the conic degenerates to a pair of
lines (i.e. we get two ${\bf CP}^1$s instead of just one). The pairs
of lines intersect $X_{n-1}$ in a `spectral divisor' $C_{n-2}$. This
divisor is automatically compatible with the ${\bf Z}_2$ involution
of $X_{n-1}$.

The appearance of a conic bundle is familiar from the geometric
engineering of gauge groups of type $A_n$ or $D_n$. Indeed, ALE
spaces of type $A_n$ or $D_n$ can both be thought of as affine
conic bundles. (For exceptional gauge groups, we need elliptic
fibrations). So we anticipate that $\Delta_{W_E}=0$ describes the
$D7$ locus, without even appealing to the (known) analysis of the
limiting monodromies. We will see it more explicitly later when we
use a cylinder mapping to relate modes of ${\sf C}_3$ to a
`spectral sheaf' localized at $\Delta_{W_E}=0$.

 \begin{figure}[t]
\begin{center}
            \scalebox{.7}{
               \includegraphics[width=\textwidth]{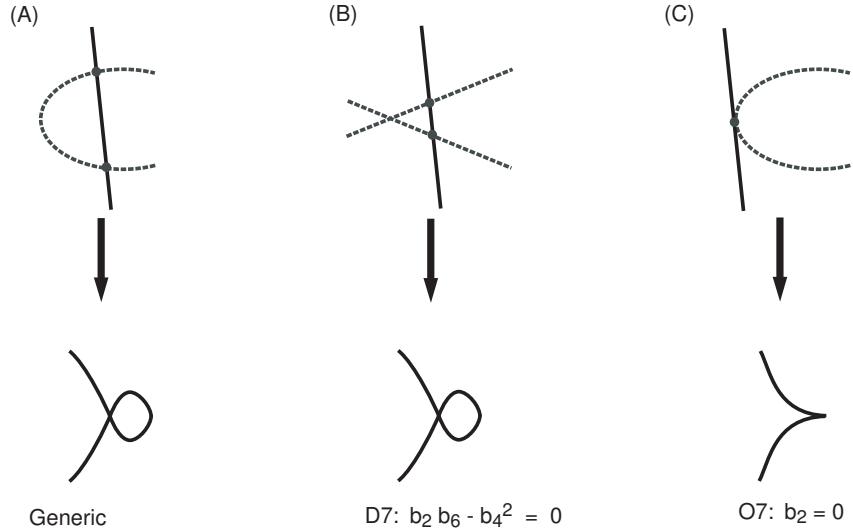}
               }
\end{center}
\vspace{-.5cm}
\begin{center}
\parbox{14cm}{\caption{ \it Picture of elliptic fibers for $g_s=0$, before and after blow-up.
(a) The generic fiber consists of a line and a conic. They intersect
in two points, with monodromy around $b_2=0$. This intersection is
identified with the IIb space-time. Contracting the conic leaves a
nodal curve. (b) At the $D7$ locus $b_2 b_6 - b_4^2=0$, the conic
degenerates to a pair of lines. Blowing down the degenerate conic
yields again a nodal curve. (c) At the $O7$ locus $b_2 = 0$ the line
and the conic are tangent. Upon blowing down the conic, we get a
cusp.}\label{SenFibers}}
\end{center}
 \end{figure}
The situation is described in pictures in figure \ref{SenFibers}.
In the limit $t \to 0$, the generic fiber is nodal, and we get
cusps over ${\sf b}_2 = 0$. The blow up replaces the generic nodal
fiber by two intersecting ${\bf P}^1$'s: a line given in equation
(\ref{LineEq}), and a conic given in equation (\ref{ConicEq}). The
blow-up separates the double point of the nodal curve into the two
intersection points of the line with the conic. The exceptional
${\bf P}^1$ (the conic) further degenerates to a pair of lines over
${\sf b}_2 {\sf b}_6 - {\sf b}_4^2 = 0$. And the cusp at ${\sf
b}_2=0$ is replaced by a line and a conic which are tangent, i.e
the two intersection points of the line and the conic in the
generic fiber coincide here.

Let us consider compactification to eight dimensions. Then $Y$ is
$K3$, $X_1$ is $T^2$ and $C_0$ corresponds to $2\times 16 = 32$
points on the $T^2$, interchanged by the involution. In other
words, we get an $SO(32)$ spectral cover on $X_1$ associated to the
vector representation, and its Fourier-Mukai transform is a (highly
reducible) $SO(32)$ vector bundle on $X_1$ of rank $32$. One may
view this as a type I compactification, or as a heterotic
compactification `with vector structure,' in the language of
\cite{Witten:1997bs}.

We can get enhanced gauge symmetry by making $W_E$ more singular.
This is particularly clear if we consider compactification to eight
dimensions. To understand this, it is perhaps useful to relate our
picture to some other constructions in the literature.

\newsubsection{Rational surfaces and $G$-bundles on an elliptic curve}

The picture we obtained is closely related to another construction
in the literature. Flat $G$-bundles over an elliptic curve can be
related to rational surfaces of type $G$. When $G = E_k$, we get
the del Pezzo surface $dP_k$ of degree $9-k$, and the elliptic
curve is an anti-canonical curve in $dP_k$. The generalization for
$G = SO(32)$ was discussed in \cite{Clingher:2003ui} and for
general gauge groups it was discussed in \cite{LeungADE,LZ1,LZ2}.

The idea is roughly as follows. Suppose we are given a surface $S$
for which the group of line bundles ${\rm Pic}(S)$ is discrete, i.e.
${\rm Pic}(S) \cong H^2(S,{\bf Z})$, like for a rational surface.
Suppose we are also given a lattice $N$, and define
\be T\ =\ N \otimes {\bf C}^* \ee
We would like to define a $T$-bundle on $S$. To each such
$T$-bundle, we can associate a line bundle (i.e. a ${\bf
C}^*$-bundle), by picking a character $\chi \in \Lambda_{char}
=\Hom(T,{\bf C}^*)$. Furthermore, ${\bf C}^*$-bundles are classified
by ${\rm Pic}(S)$, so from our $T$-bundle we get an element of
\be \Hom(\Lambda_{char}, {\rm Pic}(S)) \ee
Conversely, as $T$ is abelian, such a map reconstructs a unique
$T$-bundle. Our $T$-bundle splits as a sum of line bundles, and the
only $T$-bundle which gets mapped to zero is the $T$-bundle whose
summands are line bundles which all have their first Chern class
identically zero. So $T$-bundles on $S$ are classified by
$\Hom(\Lambda_{char}, {\rm Pic}(S))$. Further note that
\be
N^\vee \ = \ \Hom(N,{\bf Z}) \ = \ \Hom(T,{\bf C}^*)
\ee
so we can also say that $T$-bundles on $S$ are classified by
$\Hom(N^\vee,{\rm Pic}(S))$.

Now suppose that $H^2(S,{\bf Z})$ has a sublattice which is
isomorphic to a root lattice $\Lambda_{rt}$ for a Lie group $G$.
Now we also take $N = \Lambda_{wt}$ so that $N^\vee =
\Lambda_{root}$. Then, we get a canonical element of $\Hom(N^\vee,
{\rm Pic}(S))$, and therefore we get a canonical $T$-bundle on $S$.
More precisely, the map is canonical up to an automorphism of
$\Lambda_{root}$, which is given by the Weyl group. But $T$-bundles
related by an action of the Weyl group determine the same
$G$-bundle, where we identify $T$ with a maximal torus of $G$. So
we get a canonical $G$-bundle on $S$.

If our rational surface $S$ contains an elliptic curve $E$, then we
can restrict our canonical $G$-bundle on $S$ to get a $G$-bundle on
$E$. If furthermore $[E]\in \Lambda_{rt}^\perp \subset Pic(S)$ in
the sense that $\alpha(E) = 0$ for any $\alpha \in \Lambda_{rt}$,
then we get a flat $G$-bundle on $E$. As shown in
\cite{LeungADE,LZ1,LZ2}, all flat $G$-bundles on $E$ may be
recovered in this way, and moreover there exists an essentially
unique rational surface $S_G$ such that the moduli space of the flat
$G$-bundle on $E$ equals the complex structure moduli space of $S$
keeping $E$ fixed. The surface $S_G$ is rational and can be
constructed very explicitly. We will refer to $S_G$ as the rational
surface of type $G$. For $G = E_k$, one recovers the del Pezzo
surfaces.

By considering configurations of lines, we can also construct
associated bundles ${\cal V}_\rho$ for each representation $\rho$ of
$G$. By restriction, they yield associated bundles on $E$.

The main case of interest in this paper is type $D_n$, so let us
spell out the relevant surfaces. We start with a Hirzebruch surface
${\bf F}_1$. We have $H^2({\bf F}_1,{\bf Z}) = \vev{ b,f }$ with
$b^2 = -1$, $f^2 = 0$, and $b\cdot f = 1$. Now we blow up $n$ points
$l_1, \ldots, l_n$ in general position to get $S_{D_n}$. The
canonical bundle is given by $K = -2b -3f + \sum_i l_i$. The root
lattice is given by
\be \Lambda_{rt} \ = \ \{ x \in {\rm Pic}(S)\, |\ x\cdot K = x
\cdot f = 0 \} \ee
Indeed, we may take the simple roots to be given by
\be \alpha_1 = f-l_1-l_2,\quad \alpha_2 = l_1 - l_2, \quad \ldots,
\quad \alpha_n = l_{n-1} - l_n \ee

Now we consider the elliptic curve $E$ with identity $p_0 \in E$
representing the anti-canonical class. To this end, we first embed
$E$ as an anti-canonical class in ${\bf P}^2$, using the linear
system $|3 p_0|$. Next, we blow-up the image of $p_0$ in ${\bf P}^2$
to get $E$ as an anti-canonical curve in ${\bf F}_1$. Finally, we
blow-up $n$ additional points $p_i$ on $E$, or rather their image in
${\bf F}_1$, to get $E$ as an anti-canonical curve in $S_{D_n}$.
Recall that $-K \cdot l_i = 1$, so the intersections by $l_i \cap E
= p_i$ are the $n$ points on $E$ we introduced above.

We note some further aspects of this configuration. Each fiber $f$
intersects $E$ twice, and the sum of the two intersection points
is linearly equivalent to $2p_0$. So $E$ is a double covering over
$b=l_0$. It has a natural ${\bf Z}_2$ involution interchanging the
two sheets, and $p_0$ is one of the four fixed points. Also, the
involution relates each intersection point $l_i \cap E = p_i$ to
another intersection point $(f-l_i)\cap E = -p_i$. The notation
$-p_i$ is justified as $p_i + (-p_i) \sim 2p_0$, i.e. they sum to
zero in the group law on $E$.

Finally, we consider the space of pairs $(S_G,E)$, where $S_G$ is
a rational surface of type $G$ and $E \in |-K|$. Recall that for
every $v \in \Lambda_{rt}$, we get a line bundle on $S_G$, which
restricts to a line bundle $L_v$ of degree $-K \cdot v = 0$ on
$E$. Since the identity $p_0 \in E$ is given, $Pic^0(E)$ is
canonically isomorphic to $E$. So we get a natural map
\be (S_G,E) \to \Hom(\Lambda, E)/W \ee
and the image is an open dense subset. One can compactify the space
of pairs $(S_G,E)$ by including certain singular surfaces, such
that the map above extends to an isomorphism \cite{LZ1}.

Physically it is very natural that we should compactify by
including certain singular surfaces. On the heterotic side, the
moduli space has boundaries where we get an enhanced gauge
symmetry. On the $F$-theory side this should correspond to a
singular surface, whose resolution has a chain of $-2$-curves
orthogonal to the canonical class and intersecting in an ADE
pattern. At least for the $E_n$ cases one can show this is exactly
what happens.

Although here we only need the geometry of certain low degree
curves, it is very interesting to consider curves of arbitrary
degree. The correspondence predicted by heterotic/type II duality
relates BPS states obtained by quantizing membranes wrapped on a
curve $\Sigma$ in K3 with $\Sigma\cdot \Sigma = 2d-2$ to
Dabholkar-Harvey BPS states of `level $d$.' Their number is
computed by the left-moving partition function of a bosonic string
compactified using the Narain lattice $2\Gamma_{E_8} \oplus 4 H$.
We conjecture that there is an analogous correspondence for all the
$S_G$, with the Narain lattice containing the root lattice for the
group $G$.

For the case of $D_{16} = SO(32)$, $S_G$ is exactly the conic
bundle $W_E$ over ${\bf P}^1$ that we obtained in the Sen limit
from a $K3$-surface. To see this, first note that just like
$S_{SO(32)}$, $W_E$ is a ${\bf P}^1$-fibration over ${\bf P}^1$
such that the fiber splits into a pair of lines $\{l_i, f - l_i\}$
for $i = 1 \ldots 16$ when $\Delta_{W_E} = 0$. In other words it is
clearly a Hirzebruch surface blown up in $16$ points, and the only
thing left to check is the self-intersection number of the base.

It is interesting to see how the general discussion of the
surfaces $S_G$ fits exactly with our expectations about the
IIb/$SO(32)$ limit. The intersection of $W_E$ with $W_T$ is given
by the curve $v=0$ in equation (\ref{ConicEq}). It is a bisection
of $W_E \to B$ with branch points over ${\sf b}_2 = 0$. This is
precisely the elliptic curve $E$ on $S_{SO(32)}$. The rank 32
bundle associated to the vector representation of $SO(32)$ is
simply given by
\be {\cal V}_{S_G} \ = \ \bigoplus_i \cO(l_i)_{S_G} \oplus
\cO(f-l_i)_{S_G} \ee
By restricting ${\cal V}$, and since $l_i \cap E = p_i$ and $(f -
l_i)\cap E = 2p_0-p_i$, we get the associated $SO(32)$ bundle on
$X_1$ given by
\be {\cal V}_E \ = \ \bigoplus_i \cO_E(p_i) \oplus \cO_E(2p_0-p_i)
\ee
By tensoring with $\cO(-p_0)$ we get the associated flat $SO(32)$
bundle:
\be {\cal V}_E\otimes \cO_E(-p_0) \ = \ \bigoplus_i \cO_E(p_i-p_0)
\oplus \cO_E(-(p_i-p_0)) \ee
Note that $\cO(l_0)|_E = \cO(p_0)$ so we could instead have started
with the bundle ${\cal V}_{S_G}\otimes \cO(-l_0)$ and restrict that
to $E$. In any case, we see that the spectral cover (aka the
$D7$-branes) precisely corresponds to the inverse image of
$\Delta_{W_E}=0$ under the projection $E \to B$. One can also
construct the associated spinor bundles, see \cite{LZ1}.

We can further consider a relative version of these
correspondences, by fibering over a base. This has been partially
worked out for the case of $G= E_k$ discussed in
\cite{Kanev,Curio:1998bva}. It seems natural to expect a
correspondence between the following categories:
\begin{enumerate}
  \item pairs $(Z_{n-1},V_{Z_{n-1}})$, where $Z_{n-1}$ is an
      elliptic Calabi-Yau with section, and $V$ is a
      holomorphic $G$-bundles on $Z_{n-1}$, semi-stable on the
      generic fiber;
  \item triples $(Z_{n-1},C_{Z_{n-1}},L_{Z_{n-1}})$ where
  $(C_{Z_{n-1}},L_{Z_{n-1}})$ is the spectral data for
  $V_{Z_{n-1}}$, i.e. $C_{Z_{n-1}}$ is the spectral cover and
  $L_{Z_{n-1}}$ is the spectral sheaf.
  \item triples $(Y_n, Z_{n-1}, [C_3])$ where $(Y_n,Z_{n-1})$
      is log Calabi-Yau and fibered by $S_G$, such that for
      each fiber we recover the dictionary between $S_G$ and
      $V|_E$ discussed above. The twisting data for this
      fibration is the Deligne cohomology class $[C_3]$ which
      lives in a certain primitive part of the cohomology of
      $Y_n$. It corresponds to the twisting data (the spectral
      sheaf $L_{Z_{n-1}})$ of $V_{Z_{n-1}}$. Further discussion
      of such Deligne cohomology classes can be found in
      section \ref{ConicCylinder}.
  \item triples $(Y_n, Z_{n-1}, V_{Y_n})$ where $(Y_n,Z_{n-1})$
      is as above, and $V_{Y_n}$ is a canonical $G$-bundle over
      $Y_n$ whose restriction to $Z_{n-1}$ yields
      $V_{Z_{n-1}}$. We expect that that the differential
      character $[\omega_3(V_{Y_n})]$, where $\omega_3$ is the
      Chern-Simons three-form and $p_1=d\omega_3$ is the first
      Pontryagin class, is equal to the Deligne cohomology
      class above up to a shift by a similar class coming from
      the log tangent bundle.
\end{enumerate}
A correspondence along these lines has been previously suggested in
\cite{FM,ChenLeung}. It would be very interesting (but require some
effort) to work this out more precisely.

At any rate, the Sen limit is more general than the $SO(32)$ limit,
since it does not require fibrations by a $K3$ surface. All we get
in general is the structure of a conic bundle. The $A_n$ and $D_n$
surfaces are both special cases of conic bundles. So rather than
investigating the above conjectural correspondences, we now move on
to study conic bundles.

\newsubsection{The cylinder map for conic bundles}

\subseclabel{ConicCylinder}

In this section we would like to establish the IIb/$F$-theory
duality map. Let us recall the main features of the central fiber.
We have
\be Y_0 \ = \ W_T \cup_{X_{n-1}} W_E \ee
and $X_{n-1}=W_T \cap W_E$ is identified with the divisor of
normal crossing singularities of $Y_n$. Furthermore, $X_{n-1} \to
B_{n-1}$ is a double cover, with branch locus (i.e. $O7$-plane
locations) given by ${\sf b}_2=0$.

For $F$-theory on a smooth Calabi-Yau $Y_t$, the physical data is
related to the cohomology groups $H^k(Y_t)$ and their Hodge
decomposition. In the limit $t \to 0$ these become the logarithmic
de Rham cohomology groups $H^k_{\rm log}(Y_0)$, where we used the
shorthand notation
\be H^k_{\rm log}(Y_0) \ = \  {\mathbb
H}^k(Y_0,\Omega_{Y_0}^\bullet(\log X_{n-1})). \ee
Since $Y_0$ fails to be smooth and complete, these cohomology
groups can be broken up into several components. This partially
mirrors the fact that on the IIb side we do not have a pure Hodge
structure either, but a division into closed string modes on
$X_{n-1}$ and open string modes associated to the $D7$-branes.

More precisely, the cohomology groups $H^k_{\rm log}(Y_0)$ carry a
natural filtration, which can be obtained as follows. On $Y_0$ we
have the short exact sequence
\be 0 \ \to \ \Omega^p_{Y_0} \ \to \ \Omega^p_{Y_0}(\log(X_{n-1}))
\ \mathop{\to}^{\rm res}\ \Omega^{p-1}_{X_{n-1}} \ \to\ 0 \ee
where ${ res}$ is the Poincar\'e residue map. This gives rise to
the long exact sequence
\be\label{ResLES} \ldots \to \ H^{k-2}(X_{n-1}) \ \to \  H^k(Y_0)
\ \to \ H^k_{\rm log}(Y_0) \ \to \ H^{k-1}(X_{n-1}) \ \to \ \ldots
\ee
where the maps respect the Hodge structure, and the coboundary map
$H^{k-2}(X_{n-1})\to H^k(Y_0)$ is a Gysin map. From this we get
the filtration
\be \W_k\ \subseteq\ \W_{k+1} = H^k_{\rm log}(Y_0) \ee
where $\W_k = H^k(Y_0)/{\rm im}(H^{k-2}(X_{n-1}))$, and
$\Gr_{k+1}$ carries a pure Hodge structure. The subspace $\W_k$
does not carry a pure Hodge structure, but we can further
decompose $H^k(Y_0)$ using the Mayer-Vietoris sequence
\be \ldots \to \ H^{k-1}(X_{n-1})\ \to \ H^k(Y_0) \ \to \ H^k(W_T)
\oplus H^k(W_E) \ \mathop{\to}^{{\sf d}^k}\ H^k(X_{n-1}) \ \to
\ldots \ee
This gives a further step in the filtration, $\W_{k-1} \subset
\W_k$, where $\W_{k-1} = {\rm coker}({\sf d}^{k-1})$, such that
$\W_{k-1}$ and $\Gr_k$ both carry a pure Hodge structure, and
$\Gr_{k-1} \cong \Gr_{k+1}$. Together these give a two-step
filtration on the cohomology $H^k_{\rm log}(Y_0)$, which is known
to agree with the monodromy weight filtration on the nearby
$H^k(Y_t)$. Eg. for $k=4$ we have
\be 0\ \subseteq \ \W_3 \ \subseteq \ \W_4 \ \subseteq \ \W_5 =
H^4_{\rm log}(Y_0) \ee
with $\Gr_3 \cong \Gr_5 \cong H^3(X_{n-1})$ and $\Gr_4 = {\rm
ker}({\sf d}^4)/{\rm Im}(H^2(X_{n-1}))$. As usual, if we are to
think of $Y$ as an $F$-theory compactification rather than an
$M$-theory compactification, then there are some restrictions on
the allowed modes. Namely we only want to keep cohomology classes
that evaluate to zero on homology classes contained in the base or
that contain the elliptic fiber. We will always assume this in the
following. We consider only the graded pieces in this subsection,
and study the filtration in more detail in the next subsection.

We can use this decomposition of $H^k_{\rm log}(Y_0)$ to compare
data on $Y_0$ to data on $X_{n-1}$. The graded pieces $\Gr_{k-1}$
and $\Gr_{k+1}$ are fairly simple to analyze. It is not hard to see
that a subset of modes of ${\sf C}_3$ reproduce the IIb fields
$B^{(2)}_{NS}$ and $C^{(2)}_{RR}$. Similarly one finds that a
subset of the complex structure deformations of $Y_0$,
corresponding to logarithmic $(n-1,1)$ forms with a pole along
$X_{n-1}$, get mapped to complex structure deformations of
$X_{n-1}$ by taking the residue.

In this subsection we want to discuss the remaining piece of the
cohomology $H^k_{\rm log}(Y_0)$, isomorphic to $\Gr_k
\cong{\ker}({\sf d}^k)/{\rm Im}(H^{k-2}(X_{n-1})$, which was
missing in \cite{Donagi:2010pd} as $W_E$ was contracted there. We
want to show that there is an equivalence of the schematic form
\be \Gr_kH^k_{\rm log}(Y_0) \ \sim \ H^{k-2}_{{\rm v},-}( C_{n-2})
\ee
at least if we restrict to modes that are allowed in $F$-theory.
Here $C_{n-2}$ is the locus in $X_{n-1}$ wrapped by the
$D7$-branes. In the process one needs to deal with certain
singularities of $C_{n-2}$, which we have analyzed only up to
codimension two. So we will assume that $n \leq 4$.

The variety $W_T$ is merely a ${\bf P}^1$-fibration all of whose
fibers are non-singular. Using the Leray sequence, its cohomology
is given by
\be H^k(W_T) \ = \ H^k(B_{n-1}) \oplus H^{k-2}(B_{n-1}) \ee
Similarly we may compute the cohomology of $W_E$ using the Leray
spectral sequence. The variety $W_E$ admits a fibration $\pi_E :W_E
\to B_{n-1}$ which is a conic bundle over $B_{n-1}$, and its fibers
may degenerate. Then on the $E_2$-page, we encounter the sheaf
cohomology groups $E_2^{k-m,m}=H^{k-m}(B_{n-1}, R^m\pi_{E*}{\bf
Z})$, and  the differential is a map $d_2:E_2^{p,q}\to
E_2^{p+2,q-1}$. Since $R^1\pi_{E*}{\bf Z}$ vanishes, the sequence
degenerates at $E_2$. We also have $R^0\pi_{E*}{\bf Z}= {\bf
Z}_{B_{n-1}}$. Now we consider the remaining groups
\be  E_2^{k-2,2}\ =\ H^k(B_{n-1}, R^2\pi_{E*}{\bf Z}) \ee
Since taking cohomology commutes with base change in the highest
degree, for any point $s$ on the base we have $R^2\pi_{E*}{\bf
Z}|_s=H^2(\pi^{-1}_E(s))\cong H_2(\pi^{-1}_E(s))$. Therefore
$R^2\pi_{E*}{\bf Z} \cong {\bf Z} \oplus {\cal R}^2_p$ where the
first factor is the class of the total fiber over a point $s \in
B_{n-1}$, and ${\cal R}^2_p$ is the remainder, which is localized
over the $D7$ locus $\Delta_{W_E}=0$. Then we can define
\be H^{k+2}_p(W_E,{\bf Z}) \ \equiv \ H^{k}(B_{n-1}, {\cal
R}^2_p)\ee
Perhaps a simpler way to say it would be that $H_p^*(W_E)$
corresponds to those classes in $H^*(W_E)$ that evaluate to zero on
homology classes that are contained in the base $B_{n-1}$ or that
contain the fiber of $\pi_E: W_E \to B_{n-1}$. The cohomology of
$W_E$ is thus given by
\be H^k(W_E) \ = \ H^k(B_{n-1}) \oplus H^{k-2}(B_{n-1}) \oplus
H^k_p(W_E)\ee

Now restricting to ${\rm ker}({\sf d}^k)$, modding out by ${\rm
Im}(H^{k-2}(X_{n-1}))$, and further restricting to allowed
$F$-theory modes (which are classes that evaluate to zero on
homology classes contained in the base or containing the whole
elliptic fiber of $Y_0$), we kill almost all the pieces of
$H^k(W_T) \oplus H^k(W_E)$. We are left with the restriction of
${\rm ker}({\sf d}^k)$ to $H^k_p(W_E)$. We would now like to give
an alternative description of $H^k_p(W_E)$ in terms of some kind of
`spectral data.' In fact, it has been known for quite a while that
the Hodge structure of a conic bundle is closely related to the
Hodge structure of its discriminant locus \cite{BeauvillePrym}. As
for del Pezzo fibrations, the isomorphism between the remaining
data can be phrased as a cylinder map. For conic bundles, the
cylinder map is particularly simple.

We introduce the following notation. We have the subvariety
$\Delta_{W_E}$ in $B_{n-1}$ given by $\Delta_{W_E} = {\sf b}_2 {\sf
b}_6 - {\sf b}_4^2 = 0$, and the subvariety
$R_{n-1}=\pi_E^*\Delta_{W_E} \subset W_E$. We denote its
normalization by $\widetilde R_{n-1}$; this is the analogue of the
cylinder \cite{Kanev,Curio:1998bva,Donagi:2008ca,BeauvillePrym}.
The conic degenerates to a pair of lines over $\Delta_{W_E}$, so
$\widetilde R_{n-1}$ consists of pairs of (unembedded) lines
fibered over $\Delta_{W_E}$, which further degenerate to a double
line when ${\sf b}_2 = {\sf b}_4 = {\sf b}_6 =0$. We have a natural
inclusion $i: \widetilde R_{n-1} \to W_E$. We also introduce
$C_{n-2} = R_{n-1} \cap X_{n-1}$. It is the pre-image of
$\Delta_{W_E}$ in $X_{n-1}$, a double cover over $\Delta_{W_E}$,
and should be thought of as the locus wrapped by the $D7$ branes
before modding out by the orientifold involution. When $n=4$ the
surface $C_{n-2}$ has double point singularities along a curve $F$,
which is the fixed locus of the involution. These singularities
degenerate further to pinch points when ${\sf b}_2 = {\sf b}_4 =
{\sf b}_6 =0$.

There is a natural map $\widetilde R_{n-1} \to C_{n-2}$ which
replaces each line by the intersection with $X_{n-1}$. However,
sometimes the two lines intersect at the same point in $X_{n-1}$,
thereby yielding only a single point on $C_{n-2}$. When $n=4$ this
yields exactly the curve $F$ on $C_{n-2}$. Thus the map $\widetilde
R_{n-1} \to C_{n-2}$ factors through
\be p_R: \widetilde R_{n-1} \to \widetilde C_{n-2}, \ee
where $\nu: \widetilde C_{n-2} \to C_{n-2}$ is obtained by
requiring that each line gets mapped to its own intersection point
with $X_{n-1}$. It is also precisely the normalization. So two
distinct lines always yield two distinct points in $\widetilde
C_{n-2}$. Note that $\widetilde C_{n-2}$ itself is therefore not a
subspace of $X_{n-1}$ (though it may be viewed as a subspace of the
blow-up of $X_{n-1}$ along the singular locus of $C_{n-2}$).

There is a diffeomorphism symmetry on $\widetilde R_{n-1}$ which
interchanges the two lines in each fiber. This descends to an
involution $\rho$ on $\widetilde C_{n-2}$, which has fixed points
only in codimension two (corresponding to a double line in
$\widetilde R_{n-1}$) and $\widetilde C_{n-2}/Z_2 = \Delta_{W_E}$.
We summarize some of the relationships in the following diagram:
\be
\begin{array}{ccc}
  \widetilde R_{n-1}\ & \mathop{\hookrightarrow}^{i} &\ W_E  \\[2mm]
  \downarrow p_R &  &  \\[2mm]
  \widetilde C_{n-2}\ &   & \downarrow  \\[2mm]
  \downarrow\quad &  & \\[2mm]
  \Delta_{W_E}\ & \hookrightarrow &\ B_{n-1}
\end{array}
\ee
We see that there is a natural map between $\widetilde C_{n-2}$
and $W_E$:
\be c \ \equiv \ i_*p_R^*: \ H^{k-2}_-(\widetilde C_{n-2})\ \to\
H^{k}_p(W_E) \ee
Its inverse (up to a constant) is given by $c^* = p_{R*}i^*$. This
map preserves the integral structure and the Hodge structure, up
to a $(1,1)$ shift in the degrees. This relates the remaining data
on the $F$-theory and IIb sides.

Let us see why $H^*_p(W_E)$ is related to the odd forms on
$\widetilde C_{n-2}$. Given a class in $H^{*}_p(W_E,{\bf Z})$, we
restrict to the cylinder and then integrate over the conic fibers.
By definition of $H_p$, the integral is equal and opposite on each
of the two lines of the fiber, so it is odd under the exchange of
the two lines. Thus the resulting class on $\widetilde C_{n-2}$ is
odd under the orientifold involution.

Finally, in order to descend to $\Gr_k$, we have to restrict to
classes in $H^k_p(W_E)$ that are in the kernel of ${\sf d}^k$. Let
$j:C_{n-2} \to X_{n-1}$ denote the embedding.  Applying the
cylinder map to relate classes $H^k_p(W_E)$ to classes in
$H^{k-2}_-(\widetilde C_{n-2})$, the resulting classes are
`vanishing' classes in the kernel of the map $(j\circ \nu)_*$. We
denote such classes by $H^{k-2}_{{\rm v}}(\widetilde C_{n-2})$.
Thus we find that
\be \Gr_kH^k_{\rm log}(Y_0)\cap H^k_p \ \cong \ H^{k-2}_{{\rm
v},-}(\widetilde C_{n-2}) \ee
as advertized. We explicitly wrote $\cap H^k_p$ in order to
emphasize that in $F$-theory we have to exclude some classes which
would otherwise be allowed, but in the remainder we will not always
state this explicitly.

Let us discuss a bit more explicitly how this relates the bosonic
fields of a $D7$-brane wrapped on $C_{n-2}$ to the degenerate
$F$-theory Calabi-Yau. As promised, this will give the geometric
engineering explanation of why $\Delta_{W_E}=0$ should be
identified with the $D7$ locus of type IIb, without appealing to
$SL(2,{\bf Z})$ monodromies (which of course gives the same
answer). The bosonic fields on a $D7$-brane consist of a complex
adjoint and a gauge field.

The complex adjoint field of the eight-dimensional gauge theory
wrapped on $C_{n-2}$ describes the deformations of the $D7$ locus.
These deformations live in $H^0_+(\widetilde
C_{n-2},\nu^*N_{C_{n-2}}) \cong H^{n-2,0}_-(\widetilde C_{n-2})$.
The switch from even to odd forms is due to the fact that the
isomorphism uses the holomorphic volume form on $X_{n-1}$, which
is odd under the orientifold involution. By the correspondence
above, this gets mapped to $H^{n-1,1}_p(W_E)$. Now $W_E$ carries a
relative holomorphic $(n,0)$-form $\Omega^{n,0}_E \in
h^{0}(W_E,K_{W_E})$, and the deformations that keep the residue
fixed are parametrized precisely by $h^{n-1,1}(W_E)$.

Similarly we can compare the data associated to the gauge field on
the $D7$-brane. The Picard group of line bundles on the $D7$-brane
is isomorphic to $H^1(\cO^*_{C_{n-2}}) $. In the present context,
we further restrict this to classes that are compatible with the
orientifold involution. It is not hard to see that the Picard
group sits in a short exact sequence
\be\label{PicardSES} 0 \ \to \ {\cal J}^1H^1_-(\widetilde C_{n-2})
\ \to H^1_-(\cO^*_{\widetilde C_{n-2}}) \ \to \ H^{1,1}_{{\bf
Z},-}(\widetilde C_{n-2}) \ \to \ 0 \ee
where 
\be {\cal J}^1H^1 \ =\ H^1_{\bf C}/F^1H^1_{\bf C} + H^1_{\bf Z}
\ee
%
is the Jacobian. The discrete part is given by the the first Chern
class $c_1(L) \in H^{1,1}_{\bf Z}(C_{n-2})$, and the Jacobian
parametrizes the continous part. When the flux vanishes, the
latter corresponds to the Wilson line moduli of the gauge field.
We ignored the half-integral shift in the flux quantization law,
which corresponds to an analogous shift on the $F$-theory side.

Now we apply the cylinder map. This maps our sequence
(\ref{PicardSES}) to
\be 0 \  \to \ {\cal J}^2H^3(W_E) \ \to {\mathbb H}^4(W_E,{\cal
D}(2))\ \to \
 H^{2,2}_{\bf Z}(W_E) \ \to \ 0
 \ee
More precisely, we need to restrict this to $H^*_p$. The discrete
part corresponds to ${\sf G}$-flux. The continous part is given by
the intermediate Jacobian
\be {\cal J}^2H^3 \ = \ H^3_{\bf C}/F^2H^3_{\bf C} + H^3_{\bf Z}
\ee
When the ${\sf G}$-flux vanishes, one may think of this as
describing periods of the three-form ${\sf C}_3$. Together, these
two pieces of data determine a Deligne cohomology class in
${\mathbb H}^4(W_E,{\cal D}(2))$, where ${\cal D}(2)$ is the
Deligne complex ${\cal D}(2) =\left\{ (2\pi i)^2{\bf Z} \to
\Omega^0_{W_E} \to \Omega^1_{W_E}\right\}$.

The appearance of Deligne cohomology is not surprising. Just as
equivalence classes of holomorphic line bundles are given by
generators of the Picard group, similarly equivalence classes of
the three-form field ${\sf C}_3$ (viewed as a gerbe) are given by
Deligne cohomology classes. The statement about holomorphic line
bundles is a special case of the latter, as $H^1(C,\cO^*)$ is
isomorphic to the Deligne cohomology group ${\mathbb H}^2(C,{\cal
D}(1))$. The $(1,1)$ shift takes this to ${\mathbb H}^4(W_E,{\cal
D}(2))$ on the $F$-theory side. There are various Deligne
cohomology groups one could consider in connection with 2-gerbes.
This particular group classifies 2-gerbes together with a kind of
holomorphic connective structure. (Note that  ${\mathbb
H}^2(C,{\cal D}(2))$ classifies line bundles together with a
holomorphic connection). Thus we see that the cylinder map solves
the problem of relating equivalence classes of the $7$-brane gauge
fields in type IIb to equivalence classes of the three-form field
in $F$-theory.

In differential geometric terms, we may think of this as follows.
The map $i_*$ is a Gysin map, which can be represented by the Thom
class $\Xi_R$ of $R$ in $W_E $ \cite{BottTu,GriffHar}. So given a
gauge field $A_\mu dx^\mu$ on $D7$, we can promote it to a
three-form of the form
\be {\sf C}_3 \ = \ {\sf A}_\mu\, dx^\mu \wedge \Xi_R \ee
on $W_E$. Similarly, given an $(n-1,0)$-form $\Phi$ on $C_{n-1}$, we
can promote it to a form on $W_E$ which parametrizes infinitesimal
complex structure deformations of $W_E$, keeping $X_{n-1}$ fixed:
\be \delta\Omega_E\ =\ \delta\Phi \wedge \Xi_R \ee
Since $R$ is a divisor, $\Xi_R$ is of type $(1,1)$. This explains
the $(1,1)$ shift in the degrees above. Conversely, the map
$p_{R*}$ appearing in $c^*$ can be interpreted as integrating over
the fibers of $p_R: \widetilde R_{n-1} \to \widetilde C_{n-2}$,
which brings down the degree by $(1,1)$. This description of the
map seems to depend on a number of choices. The formulation in
terms of equivalence classes above clarifies the map between the
invariant data.

\newsubsection{Asymptotics of the superpotential in the Sen limit}
\subseclabel{Periods}

In the previous subsections, we have described a stable version of
the Sen degeneration. We saw that there was a precise dictionary
between the central fiber of the stable degeneration and the
perturbative IIb data on its boundary. In particular, we saw how
the $7$-brane gauge fields get mapped to ${\sf G}$-flux on the
central fiber.

In this subsection, we want to consider $F$-theory
compactifications on Calabi-Yau four-folds and understand the
limiting behaviour of the flux superpotential:
\be W \ = \ {1\over 2\pi} \int_{Y_t} \Omega^{4,0} \wedge {\sf G}
\ee

We could state the problem a little more generally. Let
$\Omega^{n,0}$ denote the holomorphic volume form on $Y_n$. More
precisely, let ${\cal F}^n$ denote the line bundle over ${\sf
D}\backslash \{0\}$ with fiber $H^{n,0}(Y_t)$ for $t\not = 0$. Let
$\widetilde {\cal F}^n$ be its canonical extension over $t = 0$,
and let $\Omega^{n,0}(t)$ be a local holomorphic frame. Then much
of the interesting information about the low energy theory is
contained in the periods
\be \Pi_I \ = \ \int_{C_I} \Omega^{n,0} \ee
So we can ask for want the behaviour of the periods in the stable
degeneration limit. Depending on the value of $n$, the periods can
be interpreted as computing BPS protected masses or tensions of
wrapped branes. For definiteness we consider $n=4$, in which case
the periods can also be interpreted as computing the value of the
flux superpotential, by Poincar\'e duality.

The approach to the periods considered here is a generalization of
\cite{Clingher:2003ui}, and a number of general aspects are
explained in more detail in \cite{Clingher:2003ui,Clingher:2005rd}
and \cite{DKW}. For general aspects of Hodge theory see
\cite{GriffTranscendental,VoisinHodge,PetersSteenbrink}. A nice
intuitive description is given in sections 4.3 and 4.4 of
\cite{Denef:2008wq}. More computational approaches using
Picard-Fuchs equations have been studied in many works, see for
example
\cite{Berglund:1998ej,Alim:2009rf,Jockers:2009ti,Alim:2009bx,Jockers:2008pe,
Grimm:2009ef,Grimm:2009sy,Lerche:1998nx,Lerche:1999de}.

The question of the asymptotic behaviour of the periods in a
semi-stable degeneration limit is a classic problem in Hodge
theory. Let us briefly review some of the relevant facts. Parallel
transport of homology classes (or dually cohomology classes) around
$t = 0$ yields an automorphism of $H^k(Y_t)$, which can be
represented as a matrix $M$. By a base change, we may assume that
the monodromy is unipotent, i.e. there exists an integer $\gamma$
such that $(M-{\bf 1})^\gamma = 0$. Let $N$ be the log of the
monodromy matrix, which is then nilpotent.

In this set-up, the Schmid nilpotent orbit theorem says that the
periods have the following asymptotic form in the limit $t \to 0$:
\be \vec{\Pi}(t) \ \sim\ e^{{1\over 2\pi i}\log(t) N}\vec{\Pi}_0
\ee
The expression on the right hand side is called the nilpotent
orbit. It should be thought of as a perturbative approximation to
the periods. The vector $\vec{\Pi}_0$ is the period map for the
limiting mixed Hodge structure on $H^4(Y_t)$. Thus to find the
asymptotic form of the periods, we only need a way to derive this
limiting mixed Hodge structure.

By the work of Steenbrink \cite{Steenbrink76}, the limiting mixed
Hodge structure for a semi-stable degeneration may be read off from
the logarithmic cohomology groups of the central fiber. The Hodge
filtration is found from the decomposition
\be H^4_{\rm log}(Y_0) \ =\  \sum_{p+q=4} H^p(Y_0,
\Omega_{Y_0}^q(\log X_{3})) \ee
The monodromy weight filtration is in general a bit more
complicated to describe, but for our case we already found it in
section \ref{ConicCylinder}. It is of the form
\be 0 \ \subseteq \ {\rm im}(N)\ \subseteq\ {\rm ker}(N)\
\subseteq\  H^4_{\rm log}(Y_0) \ee
with graded pieces $\Gr_5 \cong \Gr_3 \cong {\rm coker}({\sf
d}^3)$ and $\Gr_4 \cong H^2_{{\rm v},-}(\widetilde C_{2})$. When
$H^3(W_T) = H^3(W_E) = 0$ as will usually be the case, we have
${\rm coker}({\sf d}^3)= H^3(X_{3})$

Now in order to find the asymptotic form of the superpotential, we
write the period map for the nilpotent orbit. Let $\Omega_0$ be a
generator for $F^4\cap \W_5$.  In the logarithmic description, this
is a logarithmic $(4,0)$ form on $Y_0$. The nilpotent orbit is
given by
\be \Omega_0(t) \ \equiv \ e^{{1\over 2\pi i}\log(t) N} \Omega_0 \
= \ \Omega_0 + {1\over 2\pi i} \log(t) N \Omega_0 \ee
Here we simply expanded the exponential and used that $N^2=0$.
Choose a basis $\vev{e^i,f^j,g^k}$ for $H^4(Y_t,{\bf Z})$ which is
adapted to the weight filtration. That is, $\vev{e^i}$ projects to
a basis for $\W_5/\W_4$, $\vev{f^j}$ projects to a basis for
$\W_4/\W_3$, and $\vev{g^k}$ is a basis for $\W_3$. The matrix $N$
acts as $N e^i = g^i$, $Nf^j = N g^k = 0$. Fixing an isomorphism
$H^4(Y_t,{\bf Z}) \cong {\bf Z}^{\dim H^4}$ using the basis above,
we can write the period map as
\be\label{OmtPeriod} \Omega_0(t) \ = \ e^i \int_{e_i} \Omega_0 +
f^j \int_{f_j}\Omega_0 + g^k \left({1\over 2\pi i}
\log(t)\int_{e_k} \Omega_0 + \int_{g_k}\Omega_0 \right) \ee
We can write more explicit expressions for each of the terms using
the isomorphisms of Hodge structures found in section
\ref{ConicCylinder}.

Let us consider the first term in (\ref{OmtPeriod}) above. Using
the isomorphism $F^4 \Gr_5 \cong F^3H^3(X_{3})$, we can write
\be\label{F4Gr5} \int_{e_i}\Omega_0 \ = \ \int_{d_i} \Omega^{3,0}
\ee
Here $\vev{d^i}$ is a basis for $H^3(X_{3},{\bf Z})$ which gets
mapped to the image of $\vev{e^i}$ in $\Gr_5$, and $\vev{d_i}$ is
its dual. The $(3,0)$ form should be thought of as the residue of
the logarithmic $(4,0)$ form.

Now we come to the second term in (\ref{OmtPeriod}). In order to
understand it, let us consider the Hodge structure on $\W_5/\W_3$.
We can fit it in a short exact sequence
\be\label{Gr5Gr4Extension} 0 \ \to \ \Gr_4 \ \to \ \W_5/\W_3\ \to\
\Gr_5 \ \to \ 0 \ee
We are interested in the $F^4$ part of $\W_5/\W_3$. It has a
contribution from $F^4 \Gr_4$, and another from lifting the $F^4$
part of $\Gr_5$ to $\W_5/\W_3$. We actually have $F^4 \Gr_4 = 0$
(as well as $F^4 \cap \W_4 = 0$), and the $F^4$ part of $\Gr_5$ was
described in (\ref{F4Gr5}), so we only need to describe its lift to
$\W_5/\W_3$. The lift is described by the extension class of
(\ref{Gr5Gr4Extension}). It may be written explicitly in terms of a
representing homomorphism $\psi$ \cite{CarlsonMHSExt}, which
corresponds precisely to the second term given by $f^j
\int_{f_j}\Omega_0$ in (\ref{OmtPeriod}).

In order to write a more useful expression for $\psi$, we may use
the isomorphisms for the graded pieces of the Hodge structure found
in section \ref{ConicCylinder}. Then we see that the short exact
sequence (\ref{Gr5Gr4Extension}) is very similar to an extension
sequence on $X_3$ of the form
\be\label{X3C2RelCoh} 0 \ \to \ H^2_{\rm v}( C_2)\ \to\
H^3(X_3,C_2) \ \to \ H^3(X_3) \ \to \ 0 \ee
which holds for example if $C_2$ is very ample, as we will assume.
However, the above sequence uses $H^2(C_2)$, whereas $\Gr_4$ is
related to $H^2(\widetilde C_2)$. They are certainly not
isomorphic, since $H^2(C_2)$ doesn't even carry a pure Hodge
structure. So we need the precise relation between them.

Let us recall the precise relation between $C_2$ and $\widetilde
C_2$. The surface $C_2$ is invariant under the involution, and
singular along the fixed locus, which is a curve that we will call
$F$. The normalization replaces this fixed locus by a double cover
$\tilde F \to F$, which is branched at the pinch points.

By Leray, we have $H^2(\widetilde C_2, {\bf Z}) = H^2(C_2,
\nu_*{\bf Z})$. Now we can write a short exact sequence of local
systems
\be 0 \ \to {\bf Z}_{C_2} \ \to \nu_*{\bf Z}_{\widetilde C_2} \ \to
\ {\bf L}_F \ \to \ 0 \ee
where ${\bf L}_F$ is a local system supported on $F$. This gives
the long exact sequence
\be\label{C2LocLES} \ldots \to H^1({\bf L}_F) \ \to \ H^2(C_2,{\bf
Z}) \ \to H^2(\widetilde C_2, {\bf Z}) \ \to H^2({\bf L}_F) \to
\ldots \ee
So to find the relation between $H^2(\widetilde C_2)$ and
$H^2(C_2)$, we need to know more about $H^2({\bf L}_F) $. We do
this by considering the analogous short exact sequence on $F$:
\be 0 \ \to\ {\bf Z}_{F} \ \to\ \nu_*{\bf Z}_{\widetilde F} \ \to \
{\bf L}_F \ \to \ 0 \ee
In our generic situation, $\tilde F$ is a smooth curve with a $Z_2$
involution and isolated fixed points, and $F$ is its quotient. The
associated long exact sequence is
\be\label{FLocLES}
 \ldots \to\ H^2(F,{\bf Z})\ \to\ H^2(\tilde F,{\bf Z}) \ \to\ H^2({\bf
 L}_F)\ \to\ 0
 \ee
We have that $H^2(F,{\bf Z}) = H^2(\tilde F,{\bf Z})={\bf Z}$, and
the map $H^2(F,{\bf Z}) \to H^2(\tilde F,{\bf Z})$ is
multiplication by two. Therefore we have $H^2({\bf
 L}_F)  = {\bf Z}_2$. But Hodge theory depends on the rational
structure, so we should kill the torsion. Going back to
(\ref{C2LocLES}), we see that
\be H^2(\widetilde C_2, {\bf Q})\ \cong\ H^2(C_2,{\bf Q})/{\rm
Im}(H^1({\bf L}_F)) \ee
where we denoted the rational version of ${\bf L}_F$ by the same
name. From the remaining part of the long exact sequence
(\ref{FLocLES}), we see that $H^1({\bf L}_F)\cong H^1(\tilde
F)/H^1(F)\cong H^1_-(\tilde F)$.

Now when $C_2$ is very ample, $H^2(C_2)$ decomposes as $H^2_{\rm
v}(C_2) \oplus j^*H^2(X_3)$, where $j:C_2 \to X_3$ is the inclusion
and $H^*_{\rm v} = {\rm ker}(j_*)$. So we can consider the image of
$H^1_-(\widetilde F)$ in $H^2_{\rm v}(C_2)$ by projecting. We again
denote this by ${\rm Im}(H^1_-(\widetilde F))$. Then from
(\ref{X3C2RelCoh}) we get the sequence
\be 0 \ \to \ H^2_{\rm v}( C_2)/{\rm Im}(H^1_-(\widetilde F))\ \to\
H^3(X_3,C_2)/{\rm Im}(H^1_-(\widetilde F)) \ \to \ H^3(X_3) \ \to \
0 \ee
We can further decompose this into even and odd parts under the
orientifold involution, and ${\rm Im}(H^1_-(\widetilde F))$
actually sits in the odd part, though we could have further
projected if that had not been the case. Thus we get the sequence
\be\label{ReducedRelCoh} 0 \ \to \ H^2_{{\rm v},-}( C_2)/{\rm
Im}(H^1_-(\widetilde F))\ \to\ H^3_-(X_3,C_2)/{\rm
Im}(H^1_-(\widetilde F)) \ \to \ H^3(X_3) \ \to \ 0 \ee
Now replacing $H^2_{{\rm v},-}( C_2)/{\rm Im}(H^1_-(\widetilde F))$
by $H^2_{{\rm v},-}(\widetilde C_2)\cong \Gr_4$, we see that
(\ref{ReducedRelCoh}) above is the sequence that is equivalent to
(\ref{Gr5Gr4Extension}).

We proceed to write the representing homomorphism $\psi$
\cite{CarlsonMHSExt} for the extension class of
(\ref{Gr5Gr4Extension}) or equivalently (\ref{ReducedRelCoh}).
Denote by $\vev{c^j}$ an integral basis for $H^2_{{\rm v},-}(
C_2)/{\rm Im}(H^1_-(\widetilde F))$ which maps to the image of
$\vev{f^j}$ in $\Gr_4$ under the isomorphism. The duals are cycles
$c_j \in H_2(C_2)$ that pair to zero with ${\rm
Im}(H^1_-(\widetilde F))$ and become homologically trivial when
embedded in $X_3$. We choose a set of three-chains $\Gamma_j$ in
$H_3(X_3,C_2)$ such that $\del \Gamma_j = c_j$ and such that the
image of $H^1_-(\widetilde F)$ in $H^3(X_3,C_2)$ evaluates to zero
on $\Gamma_j$. The representing homomorphism for the extension
class is then given by
\be \psi \ = \ \sum_i c^j \int_{\Gamma_j} \Omega^{3,0} \ee
In other words, we have found that
\be \int_{f_j} \Omega_0 \ = \ \int_{\Gamma_j} \Omega^{3,0} \ee
These expressions were studied in some detail in \cite{ClemensDef},
under the assumption that $C_2$ is smooth. In that case they vary
holomorphically in the moduli, and the critical locus is precisely
the Noether-Lefschetz locus. In the present case it seems that
these expressions still vary holomorphically, but the critical
locus corresponds to the Noether-Lefschetz locus for involution odd
classes on $\widetilde C_2$. As usual the chain integrals depend
only on the `endpoints' $c_j \subset C_2$, up to identifications by
the periods of $\Omega^{3,0}$. Indeed if we choose any other set of
$\tilde \Gamma_j$ such that $\del\tilde \Gamma_j = c^j$, then
$\tilde \Gamma_j - \Gamma_j$ is a closed cycle and the difference
in the integral is a period of $\Omega^{3,0}$. So modulo the
identifications, we may think of these expressions as localized on
the $D7$ locus.

An alternative approach to writing an expression for $\psi$, which
is in practice probably much simpler, would be to blow up $X_3$
along the intersection of the $D7$ and $O7$ locus. The proper
transform of $C_2$ is the normalization $\widetilde C_2$, and since
$X_3$ is smooth the Hodge structure of $X_3$ lifts to the blow-up
$\widetilde X_3$. However we wanted to demonstrate that it is in
principle possible to work only on $X_3$.

Finally we consider the last term (\ref{OmtPeriod}), given by
$g^k\int_{g_k}\Omega_0$. We may think of this as being associated
to the extension
\be 0 \ \to \ \W_3 \ \to\ \W_5 \ \to\ \W_5/\W_3 \ \to \ 0 \ee
We have $F^4\cap \W_3=0$ so the non-zero part comes entirely from
the extension class. In order to write the representing
homomorphism, we take an integral basis $d^k$ for $\W_3 \cong
H^3(X_3)$ and the dual basis $d_k$ for $H_3(X_3)$, and then lift
the $d_k$ to $\W_5^\vee$. We can represent the lifts by four-cycles
of the form $(c_{k,1},c_{k,2})$, where $c_1$ and $c_2$ are
four-chains on $W_E$ and $W_T$ with $\del c_{k,1} = -\del c_{k,2} =
d_k\in H_3(X_3)$. Then the representing homomorphism can be written
as the integral of the logarithmic $(4,0)$ form over these
four-cycles. By changing representatives, we see that up to natural
ambiguities given by periods of the form $\int_{e_i}\Omega_0$ and
$\int_{f_j}\Omega_0$ this integral depends only on the `end-points'
$d_k$. We will informally write it as
\be \int_{g_k}\Omega_0 \ = \  \int_{d_k} \Phi \ee

Altogether,  we found that the period map for the nilpotent orbit
can be written as
\be \Omega_0(t) \ = \ e^i\int_{d_i}\Omega^{3,0} +
f^j\int_{\Gamma_j} \Omega^{3,0} + g^k \left({1\over 2\pi i} \log(
t) \int_{d_k} \Omega^{3,0} + \int_{d_k} \Phi\right)\ee
In order to compare with the usual expressions in perturbative type
IIb, let us define
\be \tau \ = \ {1\over 2\pi i} \log(t) \ee
Then by using Poincar\'e duality on $X_3$, we see that the flux
superpotential $W = {1\over 2\pi}\int_{Y_t} \Omega^{4,0} \wedge
{\sf G}$ has the following asymptotic form in the limit $t\to 0$:
\be\label{IIbSuperW} W \ = \ \int_{X_3} \Omega^{3,0} \wedge {\sf
H}+\, W_{D7}\, + \int_{X_3} \Phi \wedge \widetilde H_{NS} +
\cO(e^{2\pi i \tau})\ee
where we defined ${\sf H} = H_{RR} + \tau\, \widetilde H_{NS}$, and
$W_{D7}$ is the superpotential for $D7$-branes wrapped on $C_2$
with worldvolume flux ${\sf F} \in H^2_-(C_2)/{\rm
Im}(H^1_-(\widetilde F))$. More explicitly, let us write the
worldvolume flux ${\sf F}$ of the $D7$-brane as
\be {\sf F}-{\sf F}_0\  =\   N_j\, c^j\ee
with $N_j c^j \in H^2_{{\rm v},-}(C_2)/{\rm Im}(H^1_-(\widetilde
F))$ and ${\sf F}_0 \in j^*H^2(X_3)$. Then $W_{D7}$ is defined to
be the following integral linear combination of chain integrals:
\be W_{D7}\ =\ \sum_j N_j\int_{\Gamma_j}\Omega^{3,0}.\ee
An alternative way to write it is $W_{D7}=\int_{D7}{\rm
Tr}(\phi^{2,0}\wedge ( \delb_{A_0} a + a^2 )+ \Phi^{2,0}_0 a^2)$
where $a=A-A_0$ and $\phi = \Phi - \Phi_0$, by analogy with (or
dimensional reduction from) the expression $\omega_{CS} = {\rm
Tr}(ad_{A_0}a+{2\over 3} a^3)$ for the Chern-Simons form. Again
since $C_2$ is not smooth, this should be thought of as living on
$\widetilde C_2$.

The exponential terms in (\ref{IIbSuperW}) are the corrections to
the nilpotent orbit. Since ${\rm Im}(2\pi i \tau)$ is precisely the
action of a $D(-1)$-instanton, these corrections should be
interpreted as computing $D(-1)$-instanton corrections to the
perturbative IIb superpotential.

\newpage

\newsection{Further aspects of the correspondence}

In this section, we would like to explain how a few additional
aspects of the relation between perturbative IIb and $F$-theory
can now be given a clear explanation.

\newsubsection{Tadpoles, Euler character and Chern-Schwartz-Macpherson classes}

Our approach gives a clear and conceptual way to match the Euler
character of the $F$-theory Calabi-Yau with a certain tadpole
constraint in type IIb. The tadpole that we want to consider is the
one associated to the RR four-form field $C_{(4)}$, with flux
$F_{(5)}$. In the $M$-theory description, this is the tadpole for
the six-form $C_{(6)}$, with two indices along the elliptic fiber.

The tadpole gets contributions from localized $D3$-branes,
curvature and fluxes. The constraint is given by
\be 0 \ =\ dF_{(7)}\ =\ N_{D_3} - {1\over 24} \chi(Y_t) + \vev{{\sf
G} ,{\sf G}}  \ee
where $\vev{{\sf G} ,{\sf G}} = \half \int_Y {\sf G} \wedge {\sf
G}$. On the other hand, in perturbative IIb we have an analogous
relation of the form
\be 0 \ = \ dF_{(5)} \ = \ N_{D_3} -{1\over 24}\chi_{\rm IIb}(O7,D7) +
{1\over {\rm im}(\tau)}\vev{{\sf H},{\sf H}} + \vev{{\sf F},{\sf F}} \ee
Here $\chi_{\rm IIb}(O7,D7)$ denotes curvature contributions from
the $D7$ and $O7$-planes, and the flux contributions are given by
$\vev{{\sf H},{\sf H}} =\int_Z{\sf H} \wedge \overline{\sf H}$ and
$\vev{{\sf F},{\sf F}}= \half \int_{D7}{\sf F} \wedge {\sf F}$. We
can try to compare the individual contribution from localized
branes, curvature and fluxes. This leads us to the expectation
that
\be \chi(Y_t)\ \to\ \chi_{\rm IIb}(O7,D7), \qquad \vev{{\sf G},{\sf
G}}\ \to\ {1\over {\rm im}(\tau)}\vev{{\sf H},{\sf H}} + \vev{{\sf
F},{\sf F}}    \ee
in the Sen limit. In this subsection, we want to explain more
explicitly how the first equality comes about. The matching of the
flux contribution follows from the polarization on ${\mathbb
H}^4(Y_0, \Omega^\bullet(\log X_3))$. Using the basis $\vev{e,f,g}$
of section (\ref{Periods}), by standard results on the monodromy
weight filtration and the cylinder map we have $|(e,f,g)|^2 =
Q_3(e,g) + Q_2(f,f)$ where $Q_3$ is the polarization on
$H^3(X_3,{\bf Z})$ and $Q_2$ is the polarization on $H^2(\widetilde
C)$, so the only thing one would need to check is the
normalizations.

Previously, comparisons of Euler characters were done using
formulae similar to those of \cite{Sethi:1996es}, which relate
Chern classes of $Y$ to Chern classes on the base of the elliptic
fibration \cite{Andreas:1999ng,Aluffi:2007sx,Collinucci:2008pf}.
The basic idea is that the Euler character of a smooth elliptic
curve is zero, so the Euler character of $Y$ only gets
contributions from the discriminant locus. One can then compare the
resulting expression for finite $t$ with the expectation from the
perturbative IIb theory. We would like to do the computation of the
Euler character directly at $t = 0$. Our computation is actually
simpler, because at $t=0$ we just have ${\bf P}^1$-fibrations,
whereas for $t \not = 0$ we have to deal with elliptic fibrations.

The Calabi-Yau manifold $Y_0 = W_T \cup_{X_{n-1}} W_E$ however has
normal crossing singularities, and its topology is generally
different from the smooth fibers. However for a normal crossing
degeneration the Betti numbers of the smooth fibers agree with the
logarithmic Betti numbers of the central fiber. Thus the Euler
character we want is computed on the central fiber by
\be\label{LogEuler}  \sum_k (-1)^k \dim H^k_{\rm log}(Y_0) \ = \
\chi(W_T) + \chi(W_E) - 2\, \chi(X_{n-1}) \ee
Note that this differs from the topological Euler character of
$Y_0$, which is given by $\chi(Y_0) =  \chi(W_T) + \chi(W_E) -
\chi(X_{n-1})$. It gets an extra contribution $-\chi(X_{n-1})$ from
the logarithmic forms, as we can see from the exact sequence
(\ref{ResLES}). Clearly the comparison that we want to do would not
work if we used the ordinary Euler character.

There is an alternative way to derive the formula (\ref{LogEuler})
that uses some results from the theory of Chern-Schwartz-Macpherson
classes, which are
defined for smooth as well as singular varieties.%
\footnote{We are grateful to P. Aluffi for a very useful
correspondence on CSM classes.} Although we strictly do not need it
here, we briefly digress to explain this because it allows one to
compare more general Chern classes.

The CSM classes are obtained from a natural transformation
\be {\rm csm}_*:{\cal C}(V) \to A_*(V) \ee
on a variety $V$. Here ${\cal C}(V)$ is the category of
constructible functions, whose elements are given by finite linear
combinations $\sum m_i {\bf 1}_{D_i}$, where $m_i \in {\bf Z}$ and
${\bf 1}_{D_i}$ denotes the function which takes the value $1$ on
the closed subset $D_i \subset V$ and zero on the complement. The
category $A_*(V)$ consists of integer linear combinations of
closed subvarieties of $V$ modulo rational equivalence, i.e.
linear combinations which are divisors of a rational function are
set to zero. There is a further natural map from $A_*(V) \to
H_*(V)$, which associates to an element of $A_*(V)$ its homology
class.

The total CSM class is now defined as ${\rm csm}_*({\bf 1}_V)$,
where ${\bf 1}_V$ is the identity function on $V$. It has the
following interesting normalization property: on a smooth variety,
${\rm csm}_*({\bf 1}_V)$ agrees with the (Poincar\'e dual of the)
total Chern class of the tangent bundle. Together with naturality
under push-forwards of proper maps, this determines the
transformation ${\rm csm}_*$ uniquely. One can show that the
degree of ${\rm csm}_*({\bf 1}_V)$ always yields the topological
Euler character. This is not an analytic invariant, so it tends to
jump under semi-stable degeneration. Thus this is not exactly what
we want.

However Verdier has shown the existence of a constructible
function $\psi$, such that ${\rm csm}_*(\psi)$ is an analytic
invariant. This is referred to as `Verdier specialization'
\cite{VerdierSpec}. In particular, it does not jump under
semi-stable degeneration. So ${\rm csm}_*(\psi)$ is the natural
definition of Chern classes on our singular variety, if we want
these classes to agree with classes on the smooth fibers under a
degeneration. We suspect that these Chern classes can probably
also be formulated in terms of logarithmic forms and the log
tangent bundle.

The function $\psi$ is easy to describe. For smooth varieties of
course we have $\psi = {\bf 1}_V$. For singular varieties with
normal crossing singularities, $\psi = m$ on a component of
multiplicity $m$, and zero at any point which lies on multiple
components.

The upshot of our discussion is that the numerical invariant of
$Y_n$ which agrees with the Euler character of the generic fiber
of our family ${\cal Y}$ is given by the degree of ${\rm
csm}_*(\psi)$. Furthermore, we have $\psi = 1$ on $W_T\backslash
X_{n-1}$ and $W_E \backslash X_{n-1}$, but $\psi = 0$ on
$X_{n-1}$. In other words, we have
\be \psi\ =\ {\bf 1}_{W_T} + {\bf 1}_{W_E} - 2 \times {\bf
1}_{X_{n-1}}. \ee
Therefore we want to calculate
\be \deg({\rm csm}_*(\psi)) \ = \ \chi(W_T) + \chi(W_E) - 2\,
\chi(X_{n-1}) \ee
which is of course exactly the same formula we found above.

The computation is now easily done. Since $W_T$ is a ${\bf P}^1$-fibration
without any singular fibers, we have
\be
\chi(W_T)\ =\ \chi({\bf P}^1)\, \chi(B_{n-1})\ =\ 2\, \chi(B_{n-1}).
 \ee
Similarly, we can use the fact that $X_{n-1}$ is a double cover of
$B_{n-1}$ branched over the orientifold locus. Assuming the
orientifold locus is smooth, we clearly have
\be
\chi(X_{n-1})\ =\ 2\, \chi(B_{n-1}) - \chi(O7). \ee
The only computation that is slightly tricky is $\chi(W_E)$, as
the conics can degenerate. Assuming that the fiber over
$\Delta_{W_E}$ always consists of a pair of lines, we would have
\be \chi(W_E) \ \sim \  2\, \chi(B_{n-1}) + \chi(\Delta_{W_E})
\ee
since the Euler character of a smooth conic is $2$, but the Euler
character of a conic that has degenerated to a pair of lines is
$3$. However the conic could further degenerate in higher
codimension, and the type of singularity depends on the dimension
$n$. To be definite, we concentrate on $F$-theory
compactifications on Calabi-Yau four-folds, i.e. we take $n=4$.
Then we only need to consider singularities up to codimension
three.

From equation (\ref{ConicEq}) for $W_E$, we then see that  when
${\sf b}_2 = {\sf b}_4 = {\sf b}_6=0$ the conic can further
degenerate to a double line given by $y^2=0$, whose topological
Euler character is $2$. These are exactly the ordinary double
point singularities of $\Delta_{W_E} = 0$. Let us denote the
number of such points by $n_d$. Then we have
\be \chi(W_E) \ = \  2\, \chi(B_{n-1}) + \chi(\Delta_{W_E}) - n_d
\ee
Adding up the contributions, the various $\chi(B_{n-1})$ all cancel, and we find
\be \deg({\rm csm}(\psi)) \ = \ 2\, \chi(O7) + \chi(\Delta_{W_E})
- n_d \ee
Now we can compare this with perturbative IIb. The curvature
contribution to the $D3$ tadpole which has been proposed in the IIb
context is \cite{Aluffi:2007sx,Collinucci:2008pf}:
\be \chi_{\rm IIb} \ = \ 2 \chi(O7) + \half \chi_0(D7) \ee
Here $\chi_0$ is defined as $\chi_0 = \chi(\widetilde C) - n_d$,
where $\widetilde C$ is a two-fold covering of $\Delta_{W_E}=0$
which two-to-one generically and one-to-one at the ordinary double
points. It follows that $\chi_0 = (2\chi(\Delta_{W_E}) -n_d)-n_d$.
Plugging in, we see that the $F$-theory expression naturally
matches with the answer expected from perturbative type IIb.

Note that the appearance of the cover $\widetilde C \to
\Delta_{W_E}$ is natural from several points of view. From the
$F$-theory perspective, the covering $\widetilde C \to
\Delta_{W_E}$ is two-to-one precisely when the fiber of the
cylinder $\widetilde R_3\to \Delta_{W_E}$ consists of a pair of
lines, and one-to-one when the fiber of $\widetilde R_3\to
\Delta_{W_E}$ is a double line (which is topologically a single
line). On the other hand it is the natural object from the point of
view of spectral covers for the vector representation of $SO(2n)$,
where it appears as the normalization of the `naive' cover. This is
why $\widetilde C$ already appeared in our $F$-theory/IIb duality
map.

\newsubsection{Singularities of $SO(2n)$ spectral covers}

We saw above that the $D7$ locus generically has singularities in
the presence of $O7$-planes. This has led to a number of puzzles
and claims about discrepancies, although many of these issues were
resolved in \cite{Collinucci:2008pf}. In this subsection, we would
like to revisit some of these issues and discuss them from the
point of view of spectral covers. We would like to emphasize here
that such singularities are in fact a well-known feature of almost
all spectral covers, including covers for the vector representation
of $SO(2n)$, or the anti-symmetric representation of the
$A_n$-series, and are completely natural. As such they were already
encountered for example in the heterotic model building literature.


Intuitively the reason for such singularities is that over a
sublocus on the base, different weights of a representation often
get mapped to the same point on the corresponding spectral cover,
for representation theoretic reasons. When that happens because two
weights are exchanged by monodromy, we expect the cover to be
smooth, but otherwise we expect singularities that can not be
gotten rid off by varying parameters. It is important to remember
however that spectral covers with such singularities still
correspond to smooth non-abelian configurations; the singularities
only appear because we insist on giving an abelianized description.

For the purpose of this paper, we are interested in the spectral
cover $C$ for the $2n$-dimensional vector representation of
$SO(2n)$, see eg. \cite{HitchinInt,DonagiCovers}. It is given by
the equation
\be\label{SOPoly} P_{SO(2n)} \ = \ \det(\lambda I - \Phi) \ = \
\lambda^{2n} + a_2 \lambda^{2n-2} + \ldots + a_{2n} \ = \ 0 \ee
i.e we only have even terms and the equation is invariant under $
\lambda \to -\lambda$. The $a_i$ are various Casimirs of the Higgs
field $\Phi$. Furthermore, $a_{2n}$ is the determinant of an
anti-symmetric matrix, so it is a square, namely the square of the
Pfaffian of $\Phi$.

This cover is singular at the fixed points of the involution
$\rho(\lambda) = -\lambda$. Near the singularities of
(\ref{SOPoly}) we can write the equation of $C$ as
\be\label{PinchEqn} z \lambda^2 + w^2 \ = \ 0 \ee
where $z$ and $w$ are local coordinates such that $z\sim
\lambda^{2n-2}+\ldots + a_{2n-2}$ and $a_{2n}\sim w^2$. In
codimension one on $C$, taking $z$ constant we see that $\lambda =
w = 0$ is a double point singularity. In codimension two on $C$,
allowing $z$ to vary we recognize $\lambda = z = w = 0$ as a
cuspidal point or pinch point singularity. Again, we emphasize that
these singularities of $C$ are artefacts of the abelianization. The
corresponding non-abelian configurations are completely smooth
(provided the Higgs bundle is stable).

The usual way to deal with the singularities of $SO(2n)$ covers is
to consider the normalization $\nu:\widetilde C \to C$
\cite{HitchinInt,DonagiCovers}. In the local coordinates above, the
normalization $\widetilde C$ is explicitly given by introducing a
new coordinate $x = -w/\lambda$, and rewriting (\ref{PinchEqn}) as
$z + w^2/\lambda^2 = 0$. In other words, we have
\be x\lambda + w = 0,  \qquad x^2 + z = 0 
\ee
Locally the surface is now well-parametrized by $x$ and $\lambda$.
The map to $C$ is simply given by projecting out $x$. Under this
projection the curve $\lambda = 0$ on $\widetilde C$, which is
parametrized by $x$, is mapped to $C$ by identifying $\pm x$. The
projection is an isomorphism for $\lambda \not = 0$. The involution
$\rho$ of $C$ lifts to the involution $\tilde \rho:\lambda \to
-\lambda$, $x \to -x$ of $\widetilde C$. One easily checks that
$\widetilde C$ is smooth and the fixed points of the involution are
precisely the lifts of the pinch points of $C$.

To get a better sense of the spectral sheaf, let us consider the
following model for an adjoint $SO(2n)$ Higgs field:
\be \Phi \ =\
\left(
  \begin{array}{cccc}
    0 & a & b & 0 \\
    -a & 0 & 0 & 1 \\
    -b & 0 & 0 & 0 \\
    0 & -1 & 0 & 0 \\
  \end{array}
\right) \ee
Then $\det(\lambda I - \Phi) = \lambda^4 + (a^2 + b^2 + 1)
\lambda^2 + b^2$, so this works as a local model near the
singularities (\ref{PinchEqn}). The spectral sheaf is given by the
cokernel of $(\lambda I - \Phi)$. We find that the matrix is rank
four generically, drops to rank three along generic points on the
spectral cover, and drops to rank two along generic points on the
double curve of the spectral cover. At the pinch point it is still
rank two. So the spectral sheaf is rank one along generic points on
the spectral cover $C$, and rank two along the double curve,
including the pinch point. This means that the spectral sheaf
generically lifts to a line bundle $L$ on the normalization
$\widetilde C$, since the push-forward $\nu_*L$ of a line bundle is
rank one generically and rank two at the image of the curve
$\lambda= x^2 +z=0$ (even at the pinch point, which is scheme
theoretically a double point on this curve). The line bundle $L$
should be compatible with the orientifold involution, i.e. $\tilde
\rho^*L \cong L^\vee \otimes K_{\tilde C}$.

The above behaviour of an $SO(2n)$ spectral cover seems similar to
the behaviour for the $D7/O7$ planes derived by Sen. This is not
surprising given the relation between $SO(32)$ type I (or
heterotic) on $T^2$ and IIb with $D7/O7$ branes on $T^2$, which is
by $T$-duality on the $T^2$. By fibering the elliptic curve over a
base, it is clear that the spectral data of a type I $SO(32)$
bundle must agree exactly with the $D7/O7$ data of Sen.

We make a brief comment about the $D$-terms. Let ${\cal L}$ denote
the spectral sheaf.
Given what is known about principal $SO(2n)$ Higgs bundles, we
expect $D$-flatness to be equivalent to the stability condition
\be {\cal K} \subset {\cal L}\quad  \Rightarrow\quad \mu({\cal K})
< \mu({\cal L}) \ee
where the slope $\mu$ is defined with respect to an ample line
bundle $\cO(J)$ using the Hilbert polynomial. (We will ignore the
issue of Mumford-Takemoto stability versus the slightly stronger
condition of Gieseker stability here; they ar very similar but use
a slightly different notion of the slope). Due to the relation
$\rho^*{\cal L} \cong {\cal L}^\vee$, the slope of ${\cal L}$
vanishes automatically. So ${\cal L}$ is stable if there exist no
subsheaves of positive or zero slope. Note that the subsheaf ${\cal
K}$ is not required to be compatible with the orientifold
involution. It is crucial that we use stability and not some
primitiveness condition of the form $F\wedge J = 0$, since $D7/O7$
systems  are intrinsically non-abelian at the singularities.

\newsubsection{$D3$-instantons versus $M5$-instantons}

Another interesting issue is the comparison between $D3$-instantons
in IIb and $M5$-instantons in $F$-theory as $t \to 0$. We will
briefly review some results from
\cite{Blumenhagen:2010ja,Donagi:2010pd}, and point out some issues
that could not be resolved at the time. Given the explicit
dictionary derived here between $F$-theory data and IIb data for $t
= 0$, we can now fill some of these gaps.

The $M5$-branes in question wrap the elliptic fiber. Since the
elliptic fiber splits into two in the Sen limit, the $M5$-brane
splits into two pieces as well:
\be
M5_T = M5 \cap W_T, \qquad M5_E = M5\cap W_E
\ee
Recall also that the worldvolume of the IIb $D3$ instanton (before
orientifolding) is the intersection of the $M5$ with the normal
crossing divisor $X_{n-1}$, or equivalently it is the intersection
$D3 =M5_T \cap M5_E$.

The contribution of an $M5$ instanton to the superpotential is
given by the $M5$ partition function, after factoring out four
universal bosonic zero modes and two universal fermionic zero
modes. The worldvolume theory of the $M5$ consists of five scalars
$\phi$, a spinor $\psi$ with eight on-shell degrees of freedom, and
a chiral two-form $B^+$. Thus the partition function is of the
schematic form
\be
Z_{M5} \ =\ Z_\phi Z_\psi Z_{B^+}
\ee
In the Sen limit the $M5$ worldvolume has become reducible (with
normal crossing singularities), so we have be more careful in
saying what we mean by the partition function. Without a UV
completion we cannot derive this from first principles, but we can
give a reasonable prescription, because the singularities are
rather mild. Our point of view will be that the modes of the
worldvolume fields have to be glued along the intersection $M5_T
\cap M5_E$, analogous to the non-trivial gluings of reducible
$D$-branes discussed in
\cite{Donagi:2010pd,Donagi:2011jy,Donagi:2011dv}. Then the zero
modes of the various fields are described by the logarithmic
cohomology groups $H^k_{\rm log}(M5)$, in the same way as the zero
modes on a smooth $M5$-brane are described by the various
$H^k(M5)$.

We can again use (\ref{ResLES}) to relate $H^k_{\rm log}(M5)$ to
the ordinary cohomology. This yields a filtration
\be 0\ \subseteq\ \W_{k-1}\ \subseteq\ \W_k\ \subseteq\
\W_{k+1}=H^k_{\rm log}(M5) \ee
with $\W_{k-1} = {\rm im}(N)$ and $\W_k = {\rm ker}(N)$, much like
we saw for the $F$-theory Calabi-Yau $Y_0$ itself. We also have the
obvious Hodge filtration on $H^k_{\rm log}(M5)$. Together with a
rational structure, these give the limiting mixed Hodge structure
for the Sen limit of the $M5$-brane.

The $D3$ partition function is of the schematic form
\be
Z_{D3} \ = \ Z_\phi Z_\psi Z_F Z_{\lambda_{37}}
\ee
In the language of the present paper, it was shown in
\cite{Donagi:2010pd} that reduction of the $M5$-brane along $M5_T$
reproduces most of the expected degrees of freedom on a
$D3$-instanton in type IIb. More precisely, the forms used for
reduction on $M5_T$ in \cite{Donagi:2010pd} can be extended along
$M5_E$. This is the gluing prescription mentioned above, and yields
\be
Z_\phi \to Z_\phi^{(5)}, \qquad Z_\psi \to Z_\psi,
\qquad Z_{B^+} \rightsquigarrow Z_\phi^{(1)} Z_F
\ee
It was further argued that the chiral two-form on the $M5$-brane
should have additional modes, which reproduce the partition
function $Z_{\lambda_{37}}$ due to chiral currents on the
intersection of the $D3$-instanton with the $D7$-branes. However
the picture in \cite{Donagi:2010pd} was too singular to explicitly
check this. Our semi-stable version of the degeneration solves
this problem, and as we will now see, these modes are indeed
present and they simply come from reduction along the fibers of
$M5_E$.

To see this, let us simply restrict the cylinder map to the
$D3$-worldvolume. This yields a map
\be c: H^{i,j}_-(\widetilde \Sigma_{37}) \ \to \
H^{i+1,j+1}_p(M5_E) \ee
where $\Sigma_{37} = D3 \cap D7$, and $\nu: \widetilde \Sigma_{37}
\to \Sigma_{37}$ is its normalization. The subscript `p' stand for
primitive, i.e. we consider the cohomology classes orthogonal to
the base and the anti-canonical. The self-duality condition on the
chiral two-form means that the fluxes of interest live in
$H^{2,1}\oplus H^{0,3}$, and further restricting to $H_p$ kills any
$(0,3)$ part. Under the inverse of the cylinder map, fluxes in
$H^{2,1}$ get mapped to chiral currents $J=\del \phi$ in
$H^{1,0}(\widetilde \Sigma_{37})$.

More precisely, what we want to do is the following. Let us fix a
basis $\{A_i,B^j\}$ for $H_3(M5,{\bf Z})$ such that $A_i \cap B^j =
\delta_i^j$. We also take a basis $\omega_i$ for the imaginary
self-dual harmonic three-forms. Then up to a suitable change of
basis, we have
\be \int_{A_i} \bar\omega_j = \delta_{ij}, \qquad \int_{B^j}\bar
\omega_i = \tau_{ij} \ee
where $\tau_{ij}$ is the period matrix. We also define $z^i$ to be
periods of ${\sf C}_3$ (asssuming that the restriction ${\sf
G}|_{M5}$ is trivial in cohomology; the modification of the story
when ${\sf G}|_{M5}$ is non-zero is explained below):
\be {\sf C}_3 \ = \ 2\pi z^i \omega_i + c.c. \ee
Then the partition function for $B^+$ obtained from summing over
fluxes and holomorphic factorization is essentially proportional to
the theta function on the intermediate Jacobian of the $M5$-brane
\cite{Witten:1996hc}
\be Z_{B^+} \ \propto \ \Theta(\tau|z) \ = \ \sum \exp( \half n^i
n^j 2\pi i\tau_{ij} + 2\pi in^i z^i) \ee
So to get the asymptotic form of $Z_{B^+}$ as $t \to 0$, we should
simply substitute the periods $\tau_{ij}(t)$ and $z_i(t)$ for the
nilpotent orbit associated to the limiting mixed Hodge structure on
$H^3_{\rm log}(M5)$. The analysis is slightly complicated but in
principle it proceeds exactly like the derivation of the asymptotic
form of the superpotential in section \ref{Periods}.

Let us here simply note the main feature. From our stable version
of the degeneration, we see that for $t = 0$ the intermediate
Jacobian admits a fibration ${\cal J}_{M5} \to {\cal J}_{D3}$ with
fiber given by ${\cal J}_{\Sigma_{37}}$. Here by ${\cal
J}_{\Sigma_{37}}$ we mean the part of the Jacobian that is odd
under the orientifold involution. Then the theta function roughly
speaking factorizes as $\Theta_{D3} \Theta_{\Sigma_{37}}$; more
precisely when we fix the data on the $D3$ then we recover the
theta function of ${\cal J}_{\Sigma_{37}}$. The theta function
associated to ${\cal J}_{\Sigma_{37}}$ can be reinterpreted as
(being proportional to) a partition function of chiral fermions. In
the perturbative IIb theory, this is precisely the partition
function $Z_{\lambda_{37}}$ of chiral fermions obtained from
quantizing Euclidean D3-D7 strings. It is related to the chiral
two-form by bosonization, i.e. we have $J = \del \phi \sim
\lambda\lambda$. Similarly, the theta function $\Theta_{D3}$ is
proportional to the partition function $Z_F$ of $U(1)$ Yang-Mills
theory on the $D3$ \cite{Witten:1995gf}. It arises essentially from
the sum over worldvolume fluxes on the $D3$-instanton.

Thus altogether we have
\be Z_{B^+}\ \to\ Z_\phi^{(1)} Z_F Z_{\lambda_{37}} \ee
and therefore, modulo subtleties in properly defining the partition
functions, $Z_{M5}$ reproduces all the pieces of $Z_{D3}$ in the
Sen limit.

As emphasized in \cite{Donagi:2010pd}, it is important to note that
the intermediate Jacobian of the $M5$-brane is in general not
isomorphic to the Jacobian of any Riemann surface. Neither does it
admit a projection to a lower dimensional abelian variety, with
fibers that could be interpreted as the Jacobian of a Riemann
surface (or as a Prym). This happens only in special cases,
analyzed here and in \cite{Donagi:2010pd}, and even then only for a
piece of it as we saw above. Thus it is in general not possible,
nor is it necessary, to reinterpret $Z_{B^+}$ as a partition
function of fermions, using the $2d$ Bose-Fermi correspondence.

Now we would like to take a closer look at the vanishing behaviour.
Suppose that we have zero modes for the $\lambda_{37}$ fermions in
IIb. In this case the partition function vanishes, but we can still
get non-zero contributions to derivatives of the superpotential,
which inserts zero modes in the path integral. Let us see how the
computations in $F$-theory and IIb are related. Again, the picture
here was already argued in \cite{Donagi:2010pd}, but now we can
make it more precise. It is essentially completely analogous to the
relation between $F$-theory and the heterotic string (which is of
course not surprising given that for K3-fibrations we recovered the
$SO(32)$ degeneration).

We have
\be
Z_{\lambda_{37}}(\tau,z) \ \sim \int d\lambda \ e^{\int_{\Sigma_{37}} \lambda \delb_A   \lambda}
\ee
Here $A$ is the restriction of the $7$-brane gauge field, $z$ are
its periods (assuming $\int_{\Sigma_{37}}F=0)$, and $\tau$ describes
complex structure of $\Sigma_{37}$, which is sensitive to
deformations of the $7$-branes or the background Calabi-Yau
$X_{n-1}$.

Although the contribution to the superpotential vanishes when there
are $\lambda_{37}$ zero modes (eg. when $\int_{\Sigma_{37}} F\not =
0$), we may still get non-trivial contributions to derivatives of
the superpotential. In particular, chiral fields corresponds to
infinitesimal deformations of $\tau$ and $z$, so instanton
contributions to holomorphic couplings of such chiral fields
involve covariant derivatives of the partition function
$Z_{\lambda_{37}}(\tau,z)$ with respect to $\tau$ and $z$.
Differentiating with respect to the background fields pulls down
factors of $J \sim \lambda\lambda$ from the exponent in
$Z_{\lambda_{37}}$. These can absorb fermion zero modes and lead to
a non-vanishing path integral.

On the $M5$-side, we may not be able to recover the desired
correlators by differentiating $Z_{B^+}$, as the corresponding
chiral fields may be massive in the $M$-theory picture. However
there are still distinguished, gauge-covariant operators that we can
insert in the partition function. Suppose that our $M5$-worldvolume
contains a primitive holomorphic cycle $\alpha \in H_2(M5,{\bf Z})$.
Then we can consider the Wilson surface operator%
\footnote{In \cite{Marsano:2011nn} this was also phrased in terms
of $M2$ branes ending on the $M5$-brane. We prefer to phrase it in
terms of operators since we are not changing the solution of the
equations of motion that we are expanding around.}
\be
 W(\alpha) \ = \ e^{\int_\alpha B^+}
\ee
Note
that these operators transform non-trivially under gauge
transformations: Since $\delta B^+ = \lambda_X\, \omega^X$, we have
\be W(\alpha)\ \to\  e^{\int_\alpha B^+ + \lambda_X \omega^X}\ =\
e^{i\lambda_X Q_\alpha^X }\,W(\alpha) \ee
where $Q^X_\alpha = \int_\alpha \omega^X$. The correlation functions
of such observables inserted can in principle be computed using
holomorphic factorization. In fact with these insertion
$W(\alpha_i)$, we see that they organize as a source term for $B^+$:
\be \int_{M5} B^+ \wedge \left[{{\sf G}\over 2\pi} + \alpha_1^* +
...+ \alpha_n^*\right] \ee
The correlator is non-vanishing only if the expression in brackets
is cohomologically trivial, i.e. if it can be expressed as
$d\omega_3$. Integrating by parts, we get
\be \int_{M5} dB^+ \wedge \omega_3 \ee
so we can think of this as a shift of the periods ${\sf C}_3 \to
{\sf C}_3 + \omega_3$. If $\alpha$ shrinks to zero in the $F$-theory
limit, then this operator describes the coupling of the
$M5$-instanton to massless modes in $F$-theory, so it computes an
$M5$ instanton correction to certain derivatives of the
superpotential in $F$-theory. So by this mechanism, the $M5$
instanton may generate a contribution to a holomorphic coupling that
is forbidden in perturbation theory due to gauged $U(1)$ symmetries
\cite{Donagi:2010pd}.

The above prescription was motivated by comparison with the
bosonized description of the heterotic string, and indeed in the
degeneration limit it is easy to match this with the heterotic
string or type I string (using the spectral cover description) or
with type IIb. In the bosonized version, the currents $J$ on
$\Sigma_{37}$ come in two types. The currents that `live in the
Cartan' are of the form $J^a \sim \del \phi^a$. This clearly lifts
to an insertion of $dB^+$ in $Z_{B^+}$; it is the cylinder map we
discussed earlier. But there may exist additional currents along
$\Sigma_{37}$ of the form $J_{\pm \alpha} \sim \exp(\pm
\sqrt{2}\phi\cdot \alpha)$. Due to the way that the fields vary
over $\Sigma_{37}$ sometimes these currents are only defined at
isolated points on $\Sigma_{37}$; this point is actually crucial
for computing corrections to holomorphic couplings.

In the $M$-theory picture, these extra currents match with extra
singularities of $W_E$ in the non-generic situation, whose blow-up
yields extra cycles $\alpha \in H_2(M5,{\bf Z})$ (for example when
the $D3$-instanton intersects a stack of multiple coinciding
$D7$-branes). The heterotic/$F$-theory duality map asserts that
insertion of a current $J_\alpha$ lifts to the operator
\be
J_\alpha \ = \ e^{\sqrt{2}\phi\cdot \alpha}\ \to\ W(\alpha) \ = \ e^{\int_\alpha B^+}
\ee
We see that with this dictionary, the heterotic and $M5$-pictures
match. But formulated in this way the dictionary makes sense also
in the general case when there is no heterotic dual, i.e. if $W_E$
does not come from a $K3$-fibration.

We would like to point out in particular that although we are still
working one instanton at a time in IIb and $F$-theory, there are
already interesting results about the sum over worldsheet
instantons on the heterotic side. For example given a Calabi-Yau
three-fold $Z$, it has been found that the sum over all instantons
in any given class in $H_2(Z,{\bf Z})$ vanishes in linear sigma
model constructions \cite{Beasley:2003fx}. If we take a
$Spin(32)/Z_2$ model with vector structure and admitting an
elliptic fibration, then we can straightforwardly dualize this to a
IIb orientifold model using the Fourier-Mukai transform, without
going through $F$-theory.

The sum over worldsheet instantons becomes a sum over `vertical'
$D3$-instantons on the IIb side, i.e. instantons of the form
$\pi^{-1}_Z(C)$ where $\pi_Z: Z \to B_2$ and $C$ is a curve in
$B_2$. Horizontal $D3$-instantons wrapping the zero section of $Z$
are not included, as they map to $NS5$-instantons in the heterotic
string. Still, the result of \cite{Beasley:2003fx} shows that for
IIb orientifold duals of linear sigma model constructions, the sum
over vertical $D3$-instantons vanishes, even though the individual
contributions do not vanish. Given the state of $D3$-instanton
calculus in type IIb (or $F$-theory), this is a remarkable
statement.

\noindent {\it Acknowledgements}: It is a pleasure to thank
P.~Aluffi, S.~Katz and J.~Morgan for discussions and correspondence
related to this work. AC is supported by Simons Foundation grant
no. 208258. RD acknowledges partial support by NSF grants 0908487
and 0636606. The work of MW is supported by a Heisenberg Fellowship
from the DFG.

\end{document}